\documentclass[a4paper,11pt]{article}
\pdfoutput=1

\usepackage[svgnames]{xcolor} 

\usepackage[printonlyused,withpage]{acronym} 
\usepackage{amsfonts}
\usepackage{anyfontsize}
\usepackage{bbm}
\usepackage{booktabs} 
\usepackage{braket}
\usepackage{cancel}
\usepackage{changepage}
\usepackage{colortbl}
\usepackage{enumitem}
\usepackage[mathscr]{euscript}
\usepackage{float}
\usepackage[T1]{fontenc} 
\usepackage{gensymb}
\usepackage{Packages/jheppub}
\usepackage{listings} 
\usepackage{lmodern} 
\usepackage{makecell}
\usepackage{mathtools}
\usepackage[framemethod=tikz]{mdframed} 
\usepackage{microtype} 
\usepackage{multirow}
\usepackage{physics}
\usepackage{relsize}
\usepackage{rotating} 
\usepackage{sansmath}
\usepackage{simplewick}
\usepackage{slashed}
\usepackage{stmaryrd}
\usepackage{subfig}
\usepackage{tcolorbox}
\usepackage{theorem}
\usepackage{tikz}
\usepackage{tikz-feynman,contour}
\usepackage[normalem]{ulem}
\usepackage{xspace} 
\usepackage{yfonts}
\usepackage[vcentermath]{youngtab}

\tcbuselibrary{theorems}
\usetikzlibrary{decorations.markings}

\def\d{{\mathrm{d}}} 

\def\ii{{\text{i}}}

\makeatletter
\newcommand*{\transpose}{%
  {\mathpalette\@transpose{}}%
}
\newcommand*{\@transpose}[2]{%
  \raisebox{\depth}{$\m@th#1\intercal$}%
}
\makeatother

\newtcbox{\sln}{colback=Gainsboro,
colframe=Gainsboro}

\newcommand{\bt}[1]{{\sansmath{\boldsymbol{#1}}}}

\tikzset{snake it/.style={decorate, decoration={snake,amplitude=10mm}}}

\tikzset{/pgf/decoration/.cd,
    number of sines/.initial=10,
    angle step/.initial=20,
}

\newdimen\tmpdimen

\pgfdeclaredecoration{full sines}{initial}
{
    \state{initial}[
        width=+0pt,
        next state=move,
        persistent precomputation={
            \pgfmathparse{\pgfkeysvalueof{/pgf/decoration/angle step}}%
            \let\anglestep=\pgfmathresult%
            \let\currentangle=\pgfmathresult%
            \pgfmathsetlengthmacro{\pointsperanglestep}%
                {(\pgfdecoratedremainingdistance/\pgfkeysvalueof{/pgf/decoration/number of sines})/360*\anglestep}%
        }] {}
    \state{move}[width=+\pointsperanglestep, next state=draw]{
        \pgfpathmoveto{\pgfpointorigin}
    }
    \state{draw}[width=+\pointsperanglestep, switch if less than=1.25*\pointsperanglestep to final, 
        persistent postcomputation={
        \pgfmathparse{mod(\currentangle+\anglestep, 360)}%
        \let\currentangle=\pgfmathresult%
    }]{%
        \pgfmathsin{+\currentangle}%
        \tmpdimen=\pgfdecorationsegmentamplitude%
        \tmpdimen=\pgfmathresult\tmpdimen%
        \divide\tmpdimen by2\relax%
        \pgfpathlineto{\pgfqpoint{0pt}{\tmpdimen}}%
    }
    \state{final}{
        \ifdim\pgfdecoratedremainingdistance>0pt\relax
            \pgfpathlineto{\pgfpointdecoratedpathlast}
        \fi
   }
}

\pgfdeclaredecoration{completely sines}{initial}
{
    \state{initial}[
        width=+0pt,
        next state=upsine,
        persistent precomputation={\pgfmathsetmacro\matchinglength{
            \pgfdecoratedinputsegmentlength / int(\pgfdecoratedinputsegmentlength/\pgfdecorationsegmentlength)}
            \setlength{\pgfdecorationsegmentlength}{\matchinglength pt}
        }] {}
    \state{upsine}[width=\pgfdecorationsegmentlength,next state=downsine]{
        \pgfpathsine{\pgfpoint{0.25\pgfdecorationsegmentlength}{0.5\pgfdecorationsegmentamplitude}}
        \pgfpathcosine{\pgfpoint{0.25\pgfdecorationsegmentlength}{-0.5\pgfdecorationsegmentamplitude}}
    }
    \state{downsine}[width=\pgfdecorationsegmentlength,next state=upsine]{
        \pgfpathsine{\pgfpoint{0.25\pgfdecorationsegmentlength}{-0.5\pgfdecorationsegmentamplitude}}
        \pgfpathcosine{\pgfpoint{0.25\pgfdecorationsegmentlength}{0.5\pgfdecorationsegmentamplitude}}
}
    \state{final}{}
}

\tikzfeynmanset{ with arrow/.style = {
   decoration={
     markings,
     mark=at position 0.5
          with {\arrow[xshift=1.9mm]{Latex[width=2.75mm,length=3mm]}}
     },
   postaction=decorate}
}

\tikzfeynmanset{ with glarrow/.style = {
   decoration={
     markings,
     mark=at position 0.5
          with {\arrow[white,xshift=3mm]{Latex[width=4.125mm,length=4.5mm]}}
     },
   postaction=decorate}
}

\tikzfeynmanset{ with outarrow/.style = {
   decoration={
     markings,
     mark=at position 0.5
          with {\arrow[xshift=3.35mm]{Latex[width=4.58333mm,length=5mm]}}
     },
   postaction=decorate}
}

\tikzfeynmanset{ bigphoton/.style={
    /tikz/draw=none,
    /tikz/decoration={name=none},
    /tikz/postaction={
      /tikz/draw,
      /tikz/decoration={
        completely sines,
        segment length=3mm,
        amplitude=2.5mm,
      },
      /tikz/decorate=true,
    },
  },
}

\tikzfeynmanset{ roundbigphoton/.style={
    /tikz/draw=none,
    /tikz/decoration={name=none},
    /tikz/postaction={
      /tikz/draw,
      /tikz/decoration={
        full sines,
        segment length=3mm,
        amplitude=1.25mm,
      },
      /tikz/decorate=true,
    },
  },
}

\tikzset{ mega thick/.style= {line width = 3.4pt}
}

\hypersetup{colorlinks, breaklinks, linkcolor=black,citecolor=black,filecolor=black,urlcolor=black} 

\makeatletter
\renewcommand{\fnum@figure}{\textsc{\figurename~\thefigure}} 
\makeatother

\lstnewenvironment{pseudoc}{\lstset{frame=lines,language=C,mathescape=true}}{}
\lstnewenvironment{logs}{\lstset{frame=lines,basicstyle=\footnotesize\ttfamily,numbers=none}}{}
\lstnewenvironment{cc}{\lstset{frame=lines,language=C}}{}
\lstnewenvironment{c64}{\lstset{backgroundcolor=\color{c64},basewidth=0.65em,basicstyle=\commodoreface\color{c64light},numbers=none,framerule=10pt,rulecolor=\color{c64light},frame=tb,framexbottommargin=30pt}}{}
\lstnewenvironment{html}{\lstset{frame=lines,language=html,numbers=none}}{}
\lstnewenvironment{pseudo}{\lstset{frame=lines,mathescape=true,morekeywords={learn_string_domain, save_model}}}{}
\lstnewenvironment{pseudoctiny}{\lstset{language=C,mathescape=true,basicstyle=\tiny\sffamily}}{}
\lstnewenvironment{cctiny}{\lstset{language=C,basicstyle=\tiny\sffamily}}{}
\lstnewenvironment{pseudotiny}{\lstset{mathescape=true,basicstyle=\tiny\sffamily}}{}

\title{$\boldsymbol{t\bar{t}t\bar{t}}$ signatures through the lens of color-octet scalars}

\author{Linda M. Carpenter,}
\author{Taylor Murphy,}
\author{and Matthew J. Smylie}
\affiliation{Department of Physics, The Ohio State University\\
191 W. Woodruff Ave., Columbus, OH 43210, U.S.A.}

\emailAdd{lmc@physics.osu.edu}
\emailAdd{murphy.1573@osu.edu}
\emailAdd{smylie.8@osu.edu}

\date{\today}

\abstract{\begin{abstract}

We reinterpret two recent LHC searches for events containing four top quarks ($t\bar{t}t\bar{t}$) in the context of supersymmetric models with Dirac gauginos and color-octet scalars (\emph{sgluons}).  We explore whether sgluon contributions to the four-top production cross section $\sigma(pp \to t\bar{t}t\bar{t})$ can accommodate an excess of four-top events recently reported by the ATLAS collaboration. We also study constraints on these models from an ATLAS search for new phenomena with high jet multiplicity and significant missing transverse energy ($E_{\text{T}}^{\text{miss}}$) sensitive to signals with four top quarks. We find that these two analyses provide complementary constraints, with the jets + $E_{\text{T}}^{\text{miss}}$ search exceeding the four-top cross section measurement in sensitivity for sgluons heavier than about $800\, \text{GeV}$. We ultimately find that either a scalar or a pseudoscalar sgluon can currently fit the ATLAS excess in a range of reasonable benchmark scenarios, though a pseudoscalar in minimal Dirac gaugino models is ruled out. We finally offer sensitivity projections for these analyses at the HL-LHC, mapping the $5\sigma$ discovery potential in sgluon parameter space and computing exclusion limits at $95\%$ CL in scenarios where no excess is found.
    
\end{abstract}}

\begin{document}

\maketitle
\flushbottom

\section{Introduction}
\label{s1}

The growing collection of results from the second run of the Large Hadron Collider (LHC) offers an unprecedented opportunity to explore physics beyond the Standard Model (bSM). In addition to the steady increase in integrated luminosity during Run 2 to $\mathcal{L} \approx 140\,\text{fb}^{-1}$, advancements in multijet analysis techniques have enabled the LHC collaborations to probe rare and complex scenarios, including final states with multiple top quarks in events with large jet multiplicities and significant missing transverse energy ($E_{\text{T}}^{\text{miss}}$). For instance, a recent search \cite{ATLAS:2020new} by the ATLAS collaboration for new phenomena in four-top quark final states with at least eight jets and large $E_{\text{T}}^{\text{miss}}$ has been applied to supersymmetric (SUSY) models featuring gluinos decaying to top quarks, and may be applicable to other bSM scenarios. Meanwhile, both LHC collaborations have inched closer to a discovery at five standard deviations from the null hypothesis ($5\sigma$) of Standard Model (SM) production of four top quarks.

In the Standard Model, the process $pp\rightarrow t\bar{t}t\bar{t}$ has a cross section at next-to-leading order (NLO) in strong and electroweak couplings of $\sigma_{\text{SM}}(pp \to t\bar{t}t\bar{t}) = (12.0 \pm 2.4)\, \text{fb}$ for $pp$ collisions at a center-of-mass energy of $\sqrt{s}=13\, \text{TeV}$ \cite{SM4tNLO}. In 2020, the ATLAS collaboration reported a measurement of this cross section in final states with multiple leptons using $139\, \text{fb}^{-1}$ of $pp$ collision data from the LHC \cite{ATLAS:20204t}. ATLAS reported a cross section of $\sigma(pp \to t\bar{t}t\bar{t}) = 24^{+7}_{-6}\, \text{fb}$ with an observed significance of 4.2 standard deviations relative to the background-only hypothesis and a signal strength of $\mu = 2.0_{-0.6}^{+0.8}$ relative to the SM prediction. The significance has just been strengthened to 4.7 standard deviations above background by combination with a similar measurement in lepton(s) + jets final states \cite{ATLAS:2021rlv}. This result is interesting not only because it improves the observed significance of this rare process by more than a standard deviation; but also because the discrepancy between this measurement and the Standard Model prediction, now 1.7 standard deviations (2.0 if including \cite{ATLAS:2021rlv}), has widened from previous reports \cite{ATLAS:20184t,CMS:20194t} and exceeds that of another recently announced measurement \cite{CMS_4t_recent} by the CMS collaboration that finds no excess but carries much less significance (2.6 standard deviations) relative to background. It seems worth asking at this point which of the many bSM theories can accommodate this excess in the event that it persists or grows. Several works have appeared in the last few months that attempt to answer this question, both from an effective field theory perspective \cite{Banelli:2020iau} and in the context of specific bSM theories \cite{Hou:2020chc,Escribano:2021jne}. 

While supersymmetry remains the leading candidate for bSM physics, its most minimal and popular realizations have become ever more tightly constrained, thanks mostly to the LHC. The situation for e.g. the Minimal Supersymmetric Standard Model (MSSM) is most dire in the strong sector, with squarks excluded well into the TeV range and gluinos farther still \cite{ATLAS:2017cjl,Sirunyan_2017_1,Aaboud_2017_1,Sirunyan_2017_2,ATLAS:2017vjw,Sirunyan:2019ctn,Sirunyan:2019xwh,ATLAS-CONF-2020-002}. Extended constructions that predict phenomenology distinct from that of the MSSM are therefore increasingly well motivated. There are a plethora of ways to depart from the well trodden ground of the MSSM: one can focus on the spectrum, \`{a} la split SUSY \cite{ArkaniHamed:2004yi}; one can engineer signatures that easily evade detection at the LHC, as in the Higgsino worlds scenario \cite{Baer:2011ec}; or one can take a more top-down approach and consider alternative supersymmetry-breaking mechanisms like general gauge mediation \cite{Knapen:2015qba,Carpenter:2008he, Rajaraman:2009ga}. One framework that makes contact with all these approaches involves the imposition of a global continuous $R$ symmetry \cite{Fayet:1978qc,Hall:1991r1}, which forces all gauginos to be Dirac via a mechanism that avoids quadratically divergent contributions to scalar masses while requiring a new set of scalars that transform in the adjoint representation of each Standard Model gauge subgroup \cite{Fox:2002bu}. These Dirac gaugino models take many forms \cite{Kalinowski:2011zzc}, feature rich and unique phenomenology \cite{Dudas:2014fr, Diessner:2017sq}, are far less constrained than the MSSM \cite{Kribs:2012ss, Alvarado:2018ch, Diessner:2019sq}, and have therefore been studied in great detail \cite{Polchinski:1982an,Nelson:2002ca,Antoniadis:2006uj,Benakli:2008pg,Benakli:2009mk,Benakli:2010gi,Fok:2010vk,Kribs:2010md,Abel:2011dc,Davies:2012vu,Csaki:2013fla,Kribs:2013oda,Bertuzzo:2014bwa,Carpenter:2016lgo,Diessner:2014ksa,Fox:2014moa,diessner2015higgs,Diessner:2015yna,Diessner:2015iln,Goodsell:2015ura,diCortona:2016fsn,Braathen:2016mmb,Diessner:2016lvi,kotlarski2016analysis,Benakli:2018vqz,Liu:2019hqt}. 

A fair share of that attention has gone to the aforementioned adjoint scalars, particularly the $\mathrm{SU}(3)_{\text{c}}$ adjoint (color-octet) scalars (\emph{sgluons}) \cite{Choi:2009co,Plehn:2008ae,Chivukula:2015ef,Darme:2018rec,Carpenter:2020mrsm,Goodsell_2020}, though the weak hypercharge adjoint has been used to explain experimental anomalies in the past \cite{Carpenter:2015ucu,Benakli:2016ybe} and the weak-isospin adjoints have been scrutinized more recently \cite{Carpenter2021coloroctet,Carpenter:2021EW}. The color-octet scalars --- of which there are two, a scalar and a pseudoscalar, if they conserve CP --- couple at tree level only to colored supersymmetric particles and gluons, the latter of which allows copious pair production at the LHC. Meanwhile, these particles may receive loop-level couplings to gluon and quark pairs. In models with minimal particle content, at least, the latter coupling is proportional to the quark mass, implying that only the coupling to a $t\bar{t}$ pair is non-negligible in most cases. While sgluon decays to top quarks have been studied both at the dawn of the LHC era \cite{Choi:2009co,Plehn:2008ae} and more recently \cite{Beck_2015,Darme:2018rec,Carpenter:2020mrsm,Carpenter2021coloroctet}, the recent ATLAS measurement offers us a new opportunity to scrutinize these particles' contributions to the four-top cross section. In general, the sgluon pair production cross sections in these models are large enough that branching fractions to $t\bar{t}$ of at least $\mathcal{O}(1)\%$ should lead to detectable signatures in searches for final states with multiple top quarks.

In this work we explore how the two aforementioned recent ATLAS analyses --- the jets + $E_{\text{T}}^{\text{miss}}$ search for new phenomena \cite{ATLAS:2020new} and the multilepton final state measurement of $\sigma(pp \to t\bar{t}t\bar{t})$ \cite{ATLAS:20204t} --- favor or disfavor regions in color-octet scalar parameter space. We show that the multijet analysis has good probing power into high-mass regions of sgluon parameter space, and may currently constrain pseudoscalar sgluon scenarios in minimal ($R$-symmetric) Dirac gaugino models up to the TeV scale. We further find that scalar or pseudoscalar sgluons in scenarios with broken $R$ symmetry are good candidates to fit the measured excess in four-top events with signal strength $\mu \approx 2.0$ without being constrained by the multijet search. Sgluons heavier than $650\,\text{GeV}$ that fit this excess are currently unconstrained by decay channels not involving top quarks. Finally, we perform sensitivity studies for the full $3\,\text{ab}^{-1}$ run of the high-luminosity LHC (HL-LHC), projecting $5\sigma$ discovery limits in sgluon parameter space. We find that these two ATLAS analyses are powerful complementary discovery channels for sgluons with $t\bar{t}$ branching fractions greater than about twenty percent. We discuss how the results from these two complementary searches could aid in bSM hypothesis discrimination. We also project parameter space exclusions at $95\%$ confidence level (CL) provided no further excesses are found.

This paper is organized as follows. In \hyperref[s2]{Section 2}, we briefly review models with Dirac gauginos and color-octet scalars, and we construct an effective model of sgluon couplings to Standard Model particles, providing expressions for the effective couplings valid in minimal Dirac gaugino models. In anticipation of the following discussion, we also mention the possibility of $R$ symmetry breaking and its effect on the pseudoscalar sgluon. In \hyperref[s3]{Section 3}, we describe the two relevant ATLAS analyses in greater detail and explain how we reinterpret them within our model framework. In \hyperref[s4]{Section 4}, we provide our results and discuss realistic scenarios in which sgluons can fit the four-top excess. Here we also project the expected reach of both ATLAS analyses at the HL-LHC, showing that large regions of color-octet scalar parameter space --- including that favored by the recent cross section measurement --- can be excluded by the full planned $3\, \text{ab}^{-1}$ dataset if no statistically significant excess is observed in similar future analyses, but by the same token there is plenty of room for $5\sigma$ discovery of these color-octet scalar signatures. We summarize our work and draw conclusions in \hyperref[s5]{Section 5}.
\section{Model discussion}
\label{s2}

We begin with a brief review of models featuring Dirac gauginos and adjoint scalars, quickly pivoting to discuss the effective interactions of the latter with the Standard Model relevant to the cross section $\sigma(pp \to t\bar{t}t\bar{t})$ at the LHC. This discussion serves to establish notation and to make contact with \cite{Carpenter:2020mrsm} and \cite{Carpenter2021coloroctet} and the broader literature before proceeding with our analysis.

\subsection{Dirac gluinos and color-octet scalars}
\label{s2.1}

It is well known that Dirac gaugino masses can be generated by supersymmetry-breaking operators that introduce only finite radiative corrections. These so-called \emph{supersoft} operators \cite{Fox:2002bu},
\begin{align}\label{e1}
\mathcal{L} \supset \sum_{k=1}^3 \int \d^2 \theta\, \frac{\kappa_k}{\Lambda}\, \mathcal{W}'^{\alpha} \mathcal{W}^a_{k\alpha} \mathcal{A}_k^a + \text{H.c.},
\end{align}
consist minimally of the Standard Model gauge superfields $\mathcal{W}_k$ (with $k \in \{1,2,3\}$ such that $\mathcal{W}_3$ corresponds to $\mathrm{SU}(3)_{\text{c}}$), some hidden-sector $\mathrm{U}(1)'$ gauge superfield $\mathcal{W}'$ with nonvanishing vacuum expectation value $D'$, and a set of chiral adjoint superfields $\mathcal{A}_k$. Summation is implied in \eqref{e1} over Weyl spinor indices $\alpha$ and indices $a$ for the adjoint representation of the gauge subgroup $G_k$. The dimensionless constants $\kappa_k$, which can be unique for each $k$, parametrize the coupling of each adjoint superfield to each Standard Model gauge field. Integrating out the $\text{U}(1)'$ $D$ term from the operators \eqref{e1} yields Dirac gaugino masses. We write the mass term for the $\mathrm{SU}(3)_{\text{c}}$ ($k=3$) gaugino, the \emph{Dirac gluino} $\tilde{g}$, with spinor indices suppressed as
\begin{align}\label{e2}
\mathcal{L} \supset -m_3(\lambda^a_3 \psi^a_3 + \text{H.c.}) \equiv -m_3\, \bar{\tilde{g}}^a \tilde{g}^a\ \ \ \text{with}\ \ \ m_3 = \frac{\kappa_3}{\sqrt{2}}\frac{D'}{\Lambda}.
\end{align}
This expression explicitly shows that the Dirac gluino results from the marriage of the Majorana (MSSM-like) gluino $\lambda_3$ and the new $\mathrm{SU}(3)_{\text{c}}$ (hence color-octet) fermion $\psi_3$. Much more comprehensive reviews of these particles can be found in \cite{Carpenter:2020mrsm}.

In addition to these fermionic components, the adjoint superfields also contain scalar components, $\varphi_k$, which are complex. These scalars --- particularly the color-octet scalars, or sgluons --- have received a good deal of attention through the years, including recently \cite{Carpenter:2020mrsm,Carpenter2021coloroctet}, and they are the focus of the present work. We assume that the complex color-octet scalar $\varphi_3$ does not participate in CP-violating interactions and decompose it according to
\begin{align}\label{e3}
\varphi_3^a \equiv \frac{1}{\sqrt{2}}(O^a + \ii o^a),
\end{align}
where $O$ is a scalar (CP even) and $o$ a pseudoscalar (CP odd). We denote the mass of the scalar sgluon by $m_O$ and the mass of the pseudoscalar by $m_o$. These masses, which are in general not equal, receive contributions from multiple operators even in simple models. An unavoidable mass contribution is generated by the supersoft operator \eqref{e1} and the canonical K\"{a}hler potential for the $\mathrm{SU}(3)_{\text{c}}$ adjoint superfield $\mathcal{A}_3$ and the superfields charged under the same gauge group. The interactions between sgluons and left-chiral squarks $\tilde{q}_{\text{L}}$, for instance, originate from
\begin{align}\label{e5}
    \mathcal{L} \supset \int \d^2 \theta\, \d^2 \theta^{\dagger} \left[\mathcal{Q}^{\dagger i} \exp \left\lbrace 2 g_3 [\bt{t}_3^c \mathcal{V}_3^c]_i^{\ j}\right\rbrace \mathcal{Q}_{j} + \mathcal{A}_3^{\dagger a}\exp \left\lbrace 2 g_3 [\bt{t}^c_3 \mathcal{V}^c_3]_a^{\ b}\right\rbrace \mathcal{A}_{3b}\right],
\end{align}
where $\mathcal{Q}$ is the superfield containing left-chiral quarks $q_{\text{L}}$ and squarks $\tilde{q}_{\text{L}}$. In this expression $\bt{t}_3$ are the generators of the appropriate representations of $\mathrm{SU}(3)$ (the difference is clear from both context and indices), and $g_3$ and $\mathcal{V}_3$ are respectively the $\mathrm{SU}(3)_{\text{c}}$ running coupling and vector superfield. When the $\mathrm{SU}(3)_{\text{c}}$ $D$ term is integrated out of the sum of \eqref{e1} and \eqref{e5}, the scalar receives a mass $m_O = 2m_3$, while the pseudoscalar remains massless. It is these operators that allow the scalar to interact at tree level with squark pairs and both sgluons to interact at the same order with gluino pairs. These interactions are discussed briefly below and in great detail in \cite{Carpenter:2020mrsm}. Meanwhile, the unavoidable soft-breaking terms
\begin{align}\label{esglusoft}
\mathcal{L}_{\text{soft}} \supset -\left[2M_O^2 \tr \boldsymbol{\varphi}_3^{\dagger} \boldsymbol{\varphi}_3 + (B_O^2 \tr \boldsymbol{\varphi}_3 \boldsymbol{\varphi}_3 + \text{H.c.})\right]
\end{align}
contribute to and split the scalar and pseudoscalar masses. In this expression, the Lie-algebra valued adjoint scalar fields are decomposed according to $\boldsymbol{\varphi}_3 = \bt{t}^a_3 \varphi_3^a$, so that
\begin{align}\label{tr}
\tr \boldsymbol{\varphi}_3 \boldsymbol{\varphi}_3 = \frac{1}{2} \delta_{ab}\varphi_3^a \varphi_3^b.
\end{align}
After combining the operators \eqref{e1}, \eqref{e5}, and \eqref{esglusoft} and decomposing the adjoint scalars according to \eqref{e3}, we obtain physical sgluon masses of the form \cite{Darme:2018rec}
\begin{align}\label{phys}
\mathcal{L} \supset -\frac{1}{2}(M_O^2 + 4m_3^2+ B_O^2)O^a O^a - \frac{1}{2}(M_O^2 - B_O^2)o^a o^a.
\end{align}
But the expression \eqref{phys} is minimal and phenomenological. Additional supersymmetry-breaking operators can contribute further to the scalar and pseudoscalar masses. For example, the masses can be split by the \emph{lemon-twist} operators,
\begin{align}\label{eL}
\mathcal{L} \supset \int \d^2 \theta\, \frac{\kappa'_3}{\Lambda^2}\, \mathcal{W}'^{\alpha} \mathcal{W}_{\alpha}' \mathcal{O}^a \mathcal{O}^a,
\end{align}
which are also supersoft and cannot be forbidden by any symmetry that allows the operators \eqref{e1}.
The sgluons receive squared-mass contributions from these operators that are of $\mathcal{O}(m_3^2)$ but positive for one component and negative for the other \cite{Martin:2015eca,PhysRevD.93.075021}. This mechanism, along with the splitting term $B_O$ --- which a priori need not satisfy $B_O^2 \leq M_O^2$ --- puts one component of each adjoint in danger of becoming tachyonic. This fate can be averted using symmetry arguments \cite{Carpenter:2010rsb} or by assuming messenger-based ultraviolet completions of the supersoft operator \cite{Carpenter:2015mna}. Given the existence of myriad contributions to the sgluon mass terms, it is reasonable to view the squared masses of the scalar and pseudoscalar adjoint scalars as decoupled parameters unrestricted by \eqref{phys}. 

\subsection{$R$-symmetric sgluon interactions with Standard Model particles}
\label{s2.2}

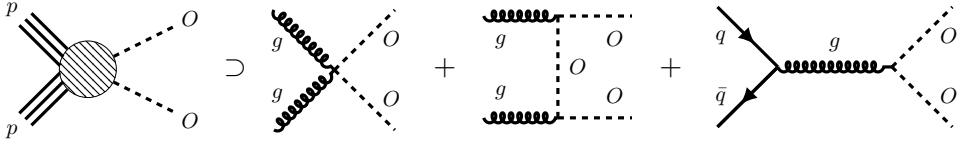
\begin{figure}\label{f1}
\begin{align*}
\scalebox{0.75}{\begin{tikzpicture}[baseline={([yshift=-.5ex]current bounding box.center)},xshift=12cm]
\begin{feynman}[large]
\vertex (i1);
\vertex [right = 1.25cm of i1, blob] (i2){};
\vertex [below left = 0.15cm of i2] (q1);
\vertex [above left=0.46 cm of q1] (r1);
\vertex [below left =0.15cm of q1] (q2);
\vertex [above left = 0.39cm of q2] (r2);
\vertex [above left=0.15cm of i2] (l1);
\vertex [below left = 0.46cm of l1] (r3);
\vertex [above left=0.15cm of l1] (l2);
\vertex [below left=0.39cm of l2] (r4);
\vertex [above=0.2cm of i2] (q3);
\vertex [above=0.2cm of q3] (q4);
\vertex [above left=1.4cm of i2] (v3);
\vertex [below left=0.15cm of v3] (p1);
\vertex [below left =0.15cm of p1] (p2);
\vertex [below left=1.4cm of i2] (v4);
\vertex [above left=0.15cm of v4] (p3);
\vertex [above left =0.15cm of p3] (p4);
\vertex [above right=0.75 cm and 1.5cm of i2] (v1);
\vertex [below right=0.75cm and 1.5cm of i2] (v2);
\diagram* {
(p1) -- [ultra thick] (r1),
(p2) -- [ultra thick] (r2),
(p3) -- [ultra thick] (r3),
(p4) -- [ultra thick] (r4),
(v3) -- [ultra thick] (i2),
(v4) -- [ultra thick] (i2),
(i2) -- [ultra thick, scalar ] (v1),
(i2) -- [ultra thick, scalar] (v2),
};
\end{feynman}
\node at (0.425,1.05) {$p$};
\node at (0.425,-1.1) {$p$};
\node at (3.55,0.9) {$O$};
\node at (3.55,-0.9) {$O$};
\end{tikzpicture}}\  \supset\  \scalebox{0.75}{\begin{tikzpicture}[baseline={([yshift=-.5ex]current bounding box.center)},xshift=12cm]
\begin{feynman}[large]
\vertex (i1);
\vertex [above left = 1.5cm of i1] (g1);
\vertex [below left = 1.5cm of i1] (g2);
\vertex [above right=1.5 cm of i1] (v1);
\vertex [below right=1.5cm of i1] (v2);
\diagram* {
(g1) -- [ultra thick, gluon] (i1),
(g2) -- [ultra thick, gluon] (i1),
(i1) -- [ultra thick, scalar] (v1),
(i1) -- [ultra thick, scalar] (v2),
};
\end{feynman}
\node at (-1,0.45) {$g$};
\node at (-1,-0.45) {$g$};
\node at (1,0.55) {$O$};
\node at (1.0,-0.5) {$O$};
\end{tikzpicture}}\ \ +\ \ \scalebox{0.75}{\begin{tikzpicture}[baseline={([yshift=-0.9ex]current bounding box.center)},xshift=12cm]
\begin{feynman}[large]
\vertex (i1);
\vertex [below = 1.8cm of i1] (i2);
\vertex [left = 1.3cm of i1] (v1);
\vertex [right= 1.3cm of i1] (v2);
\vertex [left = 1.3cm of i2] (v3);
\vertex [right=1.3cm of i2] (v4);
\diagram*{
(i2) -- [ultra thick, scalar] (i1),
(v1) -- [ultra thick, gluon] (i1),
(i1) -- [ultra thick, scalar] (v2),
(v3) -- [ultra thick, gluon] (i2),
(i2) -- [ultra thick, scalar] (v4),
};
\end{feynman}
\node at (-1,-0.45) {$g$};
\node at (-1,-1.35) {$g$};
\node at (0.35,-0.9) {$O$};
\node at (1.,-0.4) {$O$};
\node at (1,-1.4) {$O$};
\end{tikzpicture}}\ \ +\ \ \scalebox{0.75}{\begin{tikzpicture}[baseline={([yshift=-0.9ex]current bounding box.center)},xshift=12cm]
\begin{feynman}[large]
\vertex (i1);
\vertex [right = 2cm of i1] (i2);
\vertex [above left=1.5 cm of i1] (v1);
\vertex [below left=1.5cm of i1] (v2);
\vertex [above right=1.5cm of i2] (v3);
\vertex [below right=1.5cm of i2] (v4);
\diagram* {
(i1) -- [ultra thick, gluon] (i2),
(v1) -- [ultra thick, fermion] (i1),
(i1) -- [ultra thick, fermion] (v2),
(v2) -- [ultra thick] (i1),
(i2) -- [ultra thick, scalar] (v3),
(i2) -- [ultra thick, scalar] (v4),
};
\end{feynman}
\node at (-1,0.52) {$q$};
\node at (-1,-0.52) {$\bar{q}$};
\node at (1,0.4) {$g$};
\node at (3,0.55) {$O$};
\node at (3,-0.5) {$O$};
\end{tikzpicture}}
\end{align*}
\caption{Representative diagrams for scalar sgluon pair production due to gluon fusion and quark-antiquark annihilation \cite{Carpenter:2020mrsm}. Diagrams for pseudoscalar pair production are given by replacing $O \to o$ everywhere.}
\end{figure}

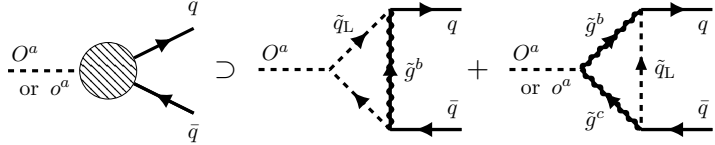
\begin{figure}\label{f2}
\begin{align*}
\scalebox{0.75}{\begin{tikzpicture}[baseline={([yshift=-.5ex]current bounding box.center)},xshift=12cm]
\begin{feynman}[large]
\vertex (i1);
\vertex [right = 1.25cm of i1, blob] (i2){};
\vertex [above right=0.75 cm and 1.5cm of i2] (v1);
\vertex [below right=0.75cm and 1.5cm of i2] (v2);
\diagram* {
(i1) -- [ultra thick, scalar] (i2),
(i2) -- [ultra thick, fermion] (v1),
(i2) -- [ultra thick ] (v2),
(v2) -- [ultra thick, fermion] (i2),
};
\end{feynman}
\node at (0.3,0.3) {$O^a$};
\node at (0.65,-0.3) {\text{or}\, $o^a$};
\node at (3.25,1.1) {$q$};
\node at (3.25,-1.1) {$\bar{q}$};
\end{tikzpicture}} \supset\ \scalebox{0.75}{\begin{tikzpicture}[baseline={([yshift=-0.75ex]current bounding box.center)},xshift=12cm]
\begin{feynman}[large]
\vertex (i1);
\vertex [right = 1.25cm of i1] (i2);
\vertex [above right=1.5 cm of i2] (v1);
\vertex [below right=1.5cm of i2] (v2);
\vertex [right=1.25cm of v1] (g1);
\vertex [right=1.25cm of v2] (g2);
\diagram* {
(i1) -- [ultra thick, scalar] (i2),
(i2) -- [ultra thick, charged scalar] (v1),
(v1) -- [ultra thick, photon] (v2),
(v2) -- [ultra thick, fermion] (v1),
(v2) -- [ultra thick, charged scalar] (i2),
(v1) -- [ultra thick, fermion] (g1),
(v2) -- [ultra thick] (g2),
(g2) -- [ultra thick, fermion] (v2),
};
\end{feynman}
\node at (0.3,0.3) {$O^a$};
\node at (3.4,0.75) {$q$};
\node at (3.4,-0.70) {$\bar{q}$};
\node at (1.5,0.8) {$\tilde{q}_{\text{L}}$};
\node at (2.73,0) {$\tilde{g}^b$};
\end{tikzpicture}}\! + \ \scalebox{0.75}{\begin{tikzpicture}[baseline={([yshift=-0.75ex]current bounding box.center)},xshift=12cm]
\begin{feynman}[large]
\vertex (i1);
\vertex [right = 1.25cm of i1] (i2);
\vertex [above right=1.5 cm of i2] (v1);
\vertex [below right=1.5cm of i2] (v2);
\vertex [right=1.25cm of v1] (g1);
\vertex [right=1.25cm of v2] (g2);
\diagram* {
(i1) -- [ultra thick, scalar] (i2),
(i2) -- [ultra thick, fermion] (v1),
(i2) -- [ultra thick, photon] (v1),
(v1) -- [ultra thick, anti charged scalar] (v2),
(v2) -- [ultra thick, fermion] (i2),
(v2) -- [ultra thick, photon] (i2),
(v1) -- [ultra thick, fermion] (g1),
(v2) -- [ultra thick] (g2),
(g2) -- [ultra thick, fermion] (v2),
};
\end{feynman}
\node at (0.3,0.3) {$O^a$};
\node at (0.65,-0.3) {\text{or}\, $o^a$};
\node at (3.4,0.75) {$q$};
\node at (3.4,-0.70) {$\bar{q}$};
\node at (1.5,0.8) {$\tilde{g}^b$};
\node at (1.5,-0.9) {$\tilde{g}^c$};
\node at (2.73,0) {$\tilde{q}_{\text{L}}$};
\end{tikzpicture}}
\end{align*}
\caption{Representative diagrams for sgluon decays to quark-antiquark ($q\bar{q}$) pairs in minimal models with Dirac gauginos \cite{Carpenter:2020mrsm}. Similar diagrams with right-handed squarks $\tilde{q}_{\text{R}}$ exist in those models. Other loops generate scalar decays to $gg$ or $g\gamma$/$gZ$.}
\end{figure}

The K\"{a}hler potential \eqref{e5} generates standard kinetic terms for the sgluons, which we write in combination with simplified sgluon mass terms as 
\begin{align}\label{e2.2.1}
\mathcal{L}_{\text{kin}} = \frac{1}{2}\left[(D_{\mu}O)^a (D^{\mu} O)^a - m_O^2\, O^a O^a\right] + \frac{1}{2}\left[(D_{\mu}o)^a (D^{\mu}o)^a - m_o^2\, o^a o^a\right],
\end{align}
where the $\mathrm{SU}(3)_{\text{c}}$-covariant derivative $D^{\mu}$ acts on sgluon fields according to
\begin{align}
    (D^{\mu}O)^a \equiv [D^{\mu}]^a_{\ \, c} O^c = (\partial^{\mu} \delta^a_{\ c} + g_3 f^{ab}_{\ \ \, c}\, g^{\mu}_{b})O^c.
\end{align}
These terms couple sgluon pairs to one or two gluons, allowing tree-level pair production at the LHC. The tree-level diagrams for these production channels are displayed in \hyperref[f1]{Figure 1}. These and subsequent diagrams were generated using the \LaTeX\ package \textsc{Tikz-Feynman} \cite{Ellis:2017fd}.
We compute these cross sections in \hyperref[s4]{Section 4}. Meanwhile, the tree-level interactions between sgluons and other colored supersymmetric particles generate interactions between the color-octet scalars and the Standard Model at one-loop order. These loop couplings have been explored at length, both in the past \cite{Plehn:2008ae,Choi:2009co} and more recently \cite{Carpenter:2020mrsm,Carpenter2021coloroctet}. Here we provide these in the minimal theory introduced above, which features a single Dirac gluino and some level of $R$ symmetry (viz. \hyperref[s2.3]{Section 2.3} or \cite{Carpenter2021coloroctet} for more on $R$ symmetry and breaking thereof). The coupling of chief interest in the present work is to pairs of top quarks. We display some of the diagrams generating this coupling in \hyperref[f2]{Figure 2}.
Integrating the squarks and gluinos out of the full theory yields an effective Lagrangian that we parametrize as
\begin{multline}\label{e2.2.2}
    \mathcal{L}_{\text{eff}} = \mathcal{L}_{\text{SM}} + \mathcal{L}_{\text{kin}} \\ + \frac{1}{4}\frac{g_3^3}{(4\pi)^2}\frac{\kappa_{g}}{\Lambda_{g}}\, d_{abc}\, O^a G_{\mu\nu}^b G^{\mu\nu\, c} + \frac{1}{2} \frac{g_1 g_3^2}{(4\pi)^2} \frac{\kappa_{B}}{\Lambda_{B}}\, O^a G^a_{\mu\nu}B^{\mu\nu}\\ + \frac{g_3^3}{(4\pi)^2}\, \sum_q m_q\, \bar{q} \bt{t}_3^a \left[\kappa_{Ot} O^a + \ii \gamma^5\, \kappa_{ot} o^a \right] q;
\end{multline}
where $\mathcal{L}_{\text{SM}}$ is the Standard Model Lagrange density. Here we have extracted from the effective couplings on the second and third lines some crucial features well known in models with Dirac gauginos, including their basic gauge coupling dependences and the linear dependence of the sgluon-to-quarks transition amplitudes on the final-state quark mass. Notice that, in this $R$-symmetric scenario, the pseudoscalar sgluon does not couple at this order to pairs of gauge bosons --- but see \hyperref[s2.2]{Section 2.2} for a discussion of alternative scenarios. These effective couplings are minimally given in terms of Passarino-Veltman functions\footnote{Our $d$-dimensional integral measure $\d^d \ell\, (2\pi)^{-d}$ differs from the measure $\d^d \ell\, (\ii \pi^{d/2})^{-1}$ frequently used elsewhere, including in the original reference.} \cite{Passarino:1979pv} by \cite{Carpenter:2020mrsm}
\begin{align}\label{e2.2.3}
\nonumber    \frac{\kappa_g}{\Lambda_g} &= 2\, \frac{m_3}{m_O^2} \times 32\ii \pi^2 \sum_{\tilde{q}} \left[m_{\tilde{q}_{\text{L}}}^2 C_0(m_O^2,0,0; m_{\tilde{q}_{\text{L}}}^2,m_{\tilde{q}_{\text{L}}}^2,m_{\tilde{q}_{\text{L}}}^2) - \{\tilde{q}_{\text{L}} \to \tilde{q}_{\text{R}}\}\right] \approx 3\,\frac{\kappa_B}{\Lambda_B},\\
\nonumber    \kappa_{Ot} &= 3 m_3 \times 16\ii \pi^2\, \frac{1}{m_O^2-4m_q^2} \left[\frac{1}{9} \mathcal{I}^{(1)}_{\tilde{q}_{\text{L}}} + \mathcal{I}_{\tilde{q}_{\text{L}}}^{(2)} - \{\tilde{q}_{\text{L}} \to \tilde{q}_{\text{R}}\}\right],\\
  \text{and}\ \ \  \kappa_{ot} &= -6m_3 \times 16\ii \pi^2 \left[ C_0(m_O^2,m_q^2,m_q^2;m_3^2,m_3^2,m_{\tilde{q}_{\text{L}}}^2) - \{\tilde{q}_{\text{L}} \to \tilde{q}_{\text{R}}\}\right],
\end{align}
where the sum in the first line is over squark flavors, and where
\begin{align*}
    \mathcal{I}_{\tilde{q}_{\text{L}}}^{(1)} &= \begin{multlined}[t][10cm] 2B_0(m_O^2;m_{\tilde{q}_{\text{L}}}^2,m_{\tilde{q}_{\text{L}}}^2) - 2B_0(m_q^2;m_3^2,m_{\tilde{q}_{\text{L}}}^2)\\ + 2(m_q^2 + m_3^2 - m_{\tilde{q}_{\text{L}}}^2) C_0(m_O^2,m_q^2,m_q^2;m_{\tilde{q}_{\text{L}}}^2,m_{\tilde{q}_{\text{L}}}^2,m_3^2)
    \end{multlined}\\
    \text{and}\ \ \ \mathcal{I}_{\tilde{q}_{\text{L}}}^{(2)} &= \begin{multlined}[t][10cm] 2B_0(m_q^2;m_3^2,m_{\tilde{q}_{\text{L}}}^2) - 2B_0(m_O^2;m_3^2,m_3^2)\\ + (2m_q^2 + 2m_3^2-2m_{\tilde{q}_{\text{L}}}^2 - m_O^2) C_0(m_O^2,m_q^2,m_q^2; m_3^2,m_3^2,m_{\tilde{q}_{\text{L}}}^2).
    \end{multlined}
\end{align*}
We note that the relation on the first line between $\kappa_g \Lambda_g^{-1}$ and $\kappa_B \Lambda_B^{-1}$ is only valid in the frequently studied scenario that all squarks except stops are degenerate and the only non-negligible loops therefore contain stops. We indeed hew to this constraint in the present work. We also (to reiterate for emphasis) assume in this work that all squarks and gluinos are heavy enough that only the decays allowed by \eqref{e2.2.2} are kinematically accessible.

\begin{figure}\label{f3}
\centering
\includegraphics[scale=0.65]{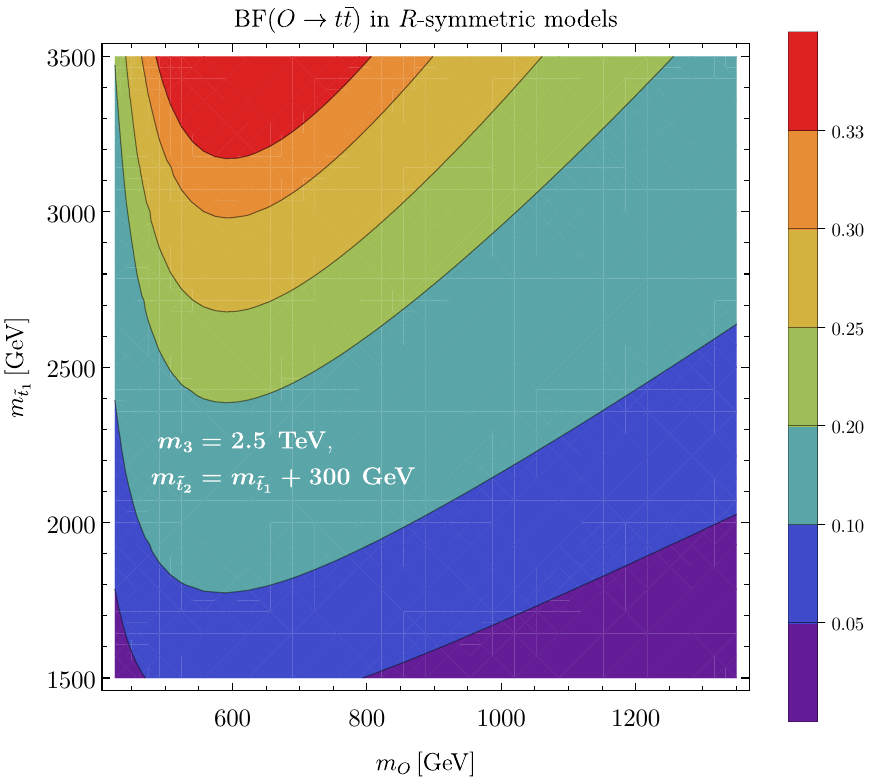}
\caption{Branching fraction $\text{BF}(O \to t\bar{t})$ of the scalar sgluon to top quark pairs in Dirac gaugino models. In this plane, the Dirac gluino mass is fixed at $m_3 = 2.5\, \text{TeV}$ and the splitting between stop squarks is $300\, \text{GeV}$. The squarks must become quite heavy for the branching fraction to exceed 40\%.}
\end{figure}

The branching fractions of the sgluons to top quark pairs strongly affect the signal strength of color-octet scalar signatures in collider events with four-top final states. In the minimal Dirac gaugino models we have discussed thus far, the pseudoscalar sgluon --- deprived of decays to Standard Model gauge bosons --- decays to top quarks with $\text{BF}(o \to t\bar{t}) \approx 1$ below the thresholds for decays to supersymmetric particles. The situation for the scalar sgluon is more complicated, as a significant portion of its decay width may be occupied by decays to pairs of gauge bosons, mostly $gg$ and $g\gamma$. The branching fraction to tops, meanwhile, is known to depend chiefly on the hierarchy between the Dirac gluino and the stop squarks \cite{Choi:2009co,Carpenter:2020mrsm}. In order to offer an idea of what scalar branching fractions are realistic in these models, we display in \hyperref[f3]{Figure 3} a contour plot of $\text{BF}(O \to t\bar{t})$ for a wide range of stop masses relative to a fixed gluino mass of $m_3 = 2.5\, \text{TeV}$. 
Here the splitting between stop squarks is fixed at $m_{\tilde{t}_2} - m_{\tilde{t}_1} = 300\, \text{GeV}$. We see that the scalar branching fraction to top quarks peaks at twice the stop mass splitting, but is universally smaller than $10\%$ for $m_{\tilde{t}_1} \lesssim 1.75\, \text{TeV}$. On the other hand, branching fractions as high as $30\%$ are attainable with $3.0\, \text{TeV}$ squarks. We use this latter scenario as a benchmark in our discussion in \hyperref[s4]{Section 4}.

\subsection{Scenarios with $R$ symmetry breaking}
\label{s2.3}

As we noted in \hyperref[s2.1]{Section 2.1}, the effective theory \eqref{e2.2.2} is an adequate description of color-octet scalar couplings to Standard Model particles in scenarios with a single Dirac gluino. Satisfying this condition conventionally requires some level of global $R$ symmetry, typically a $\mathrm{U}(1)$ symmetry that does not commute with supersymmetry, in order to forbid Majorana gluino mass terms \cite{Plehn:2008ae,Diessner:2017sq}. Enforcing such a symmetry tends to forbid a number of sgluon couplings to Standard Model particles, including that of the pseudoscalar to gluons at loop level. Introducing new Standard Model couplings at the scale in which we are interested can affect the sgluons' branching fractions to top quarks. One curious question with some practical significance to this work, therefore, is what happens if $R$ symmetry is slightly broken.

We recently conducted an investigation \cite{Carpenter2021coloroctet} of explicit $R$ symmetry breaking in models with Dirac gauginos in which we cataloged an array of interesting $R$ symmetry-breaking decays to Standard Model and supersymmetric particles. In particular, we found that a new $R$ symmetry-breaking superpotential operator including three $\mathrm{SU}(3)_{\text{c}}$-adjoint superfields generates the aforementioned pseudoscalar coupling to $gg$ via a non-vanishing gluino loop. The effective Lagrangian \eqref{e2.2.2} can be modified in this scenario to include
\begin{align}\label{e2.14}
\mathcal{L}_{\text{eff}} \supset \frac{1}{4}\frac{g_3^3}{(4\pi)^2}\frac{\kappa_{og}}{\Lambda_{og}}\, f_{c'ca'}d_{a'ab'}f_{b'bc'}\, o^a G_{\mu\nu}^b \tilde{G}^{\mu\nu\, c}\ \ \ \text{with}\ \ \ \tilde{G}^{\mu\nu\,a} = \frac{1}{2}\epsilon^{\mu\nu\alpha\beta}G_{\alpha\beta}^a.
\end{align}
The effective coupling $\kappa_{og}\Lambda_{og}^{-1}$, along with the modified effective couplings to top quarks $\kappa_{Ot}\Lambda_{Ot}^{-1}$ and $\kappa_{ot}\Lambda_{ot}^{-1}$, can be found in terms of Passarino-Veltman functions using (C.19) and (C.22) of \cite{Carpenter2021coloroctet}. (These expressions are lengthy, so we do not reproduce them here for want of space.) Suffice it to say that in Dirac gaugino models with broken $R$ symmetry, in which for instance \eqref{e2.14} does not vanish, the pseudoscalar branching fraction $\text{BF}(o \to t\bar{t})$ can be made to diminish from its $R$-symmetric value of near unity. In \cite{Carpenter2021coloroctet}, we quantified the extent of $R$ symmetry breaking with a dimensionless measure $\slashed{R}$ of $\mathcal{O}(10^{-1})$. In this scheme, the $R$-symmetric limit was given by $\slashed{R} = 0$, and e.g. $\slashed{R} = 0.10$ approximately denoted a ratio of Majorana to Dirac gluino masses of $10\%$. We found that the latter value of $\slashed{R}$ lowered the pseudoscalar branching fraction to top quarks to $40$--$60\%$. We use this scenario, and another scenario with $\slashed{R}=0.25$ --- both of which are fully explored in \cite{Carpenter2021coloroctet} --- in our discussion in \hyperref[s4]{Section 4}.
\section{Reinterpreting two ATLAS studies for color-octet scalars}
\label{s3}

The centerpiece of our investigation is a pair of analyses performed by the ATLAS collaboration using the full LHC Run 2 dataset of $139\, \text{fb}^{-1}$ of $pp$ collisions at $\sqrt{s}=13\, \text{TeV}$. The first of these is the measurement \cite{ATLAS:20204t} of the four-top production cross section in final states with multiple leptons, which provides the impetus for this work. The second, as we mentioned in the \hyperref[s1]{Introduction}, is a search \cite{ATLAS:2020new} for new phenomena in final states with many jets and significant missing transverse energy that shows sensitivity to events including four top quarks. Our goal is to understand if and how these two searches are related and how they each constrain color-octet scalars in the models discussed in \hyperref[s2]{Section 2}. To this end, we have reimplemented these analyses in version 1.9.20 of \textsc{MadAnalysis\,5} \cite{Conte_2013}, which provides a user-friendly framework to apply (in principle) any LHC analysis to a wide range of bSM theories not considered by the experimental collaborations \cite{Conte_2014,Conte_2018}. In this section, we describe our reimplementations of the ATLAS analyses, detailing the selection criteria for each search, supplying some technical details, and providing some validation of our work where possible.

\subsection{ATLAS-CONF-2020-013: $t\bar{t}t\bar{t}$ production in the multilepton final state}
\label{s3.1}

We begin with the measurement by the ATLAS collaboration of the four-top production cross section $\sigma(pp \to t\bar{t}t\bar{t})$ in final states with multiple leptons. This measurement was originally announced in ATLAS-CONF-2020-013, so we refer to it as such in this rest of this work where a short identifier is helpful, such as in plots. This measurement looks for final states with either two leptons with like charge (same-sign) or at least three leptons with no charge requirement. The analysis vetoes events with same-sign electron pairs with invariant mass consistent with a low-mass hadronic resonance ($m_{ee} < 15\, \text{GeV}$) or the decay of a $Z$ boson ($m_{ee} = (m_Z \pm 10)\, \text{GeV}$), and imposes the latter cut again on opposite-sign same-flavor (OSSF) lepton pairs within events with three or more leptons. The analysis finally requires at least six jets, at least two $b$-tagged jets, and a total scalar transverse momentum,
\begin{align}\label{e3.1}
H_{\text{T}} = \sum_i \left[p_{\text{T}i}^{\text{jet}} + p_{\text{T}i}^{\text{lepton}}\right],
\end{align}
of at least $500\, \text{GeV}$. These criteria, which are summarized in Tables \hyperref[t2]{2} and \hyperref[t3]{3}, constitute a single inclusive signal region and are imposed on leptons and jets satisfying kinematic criteria listed in \hyperref[t1]{Table 1}.
\begin{table}\label{t1}
\renewcommand\arraystretch{1.4}
\centering
\begin{tabular}{c|c c c c}
\toprule
Criterion & Electrons  & Muons & Jets & $b$-tagged jets\\
\midrule
$p_{\text{T}}\,\text{[GeV]}$ & $> 28$ & $> 28$ & $> 25$ & $> 25$\\\hline
$|\eta$| & \makecell{$< 2.47$\\ $\text{and} \notin [1.37,1.52]$} & $< 2.5$ & $< 2.5$ & $<2.5$ \\\hline
\bottomrule
\end{tabular}
\caption{
Summary of preselection criteria in ATLAS-CONF-2020-013, reproduced in part from Section 3 of \cite{ATLAS:20204t}.}
\end{table}
\begin{table}\label{t2}
\renewcommand\arraystretch{1.4}
\centering
\begin{tabular}{l|ccc}
\toprule
Selection criterion & \multicolumn{3}{c}{Description}\\
\midrule
Lepton multiplicity & \multicolumn{3}{c}{\makecell{1 same-sign (SS) lepton pair\\ or $\geq 3$ leptons without charge requirement}}\\\hline
Vetoes & \multicolumn{3}{c}{\makecell{SS electon pairs with inv. mass $m_{ee} < 15\, \text{GeV}$ or $\in [81,101]\, \text{GeV}$\\ OSSF lepton pairs with inv. mass $m_{\ell \ell} \in [81,101]\, \text{GeV}$}}\\\hline\bottomrule
\end{tabular}
\caption{Summary of common selection criteria for ATLAS-CONF-2020-013, reproduced in full from Section 3 of \cite{ATLAS:20204t}.}
\end{table}
\begin{table}\label{t3}
\renewcommand\arraystretch{1.4}
\centering
\begin{tabular}{l|ccc}
\toprule
Signal region & $N_{\text{jet}}$  & $N_{b\text{-jet}}$ & $H_{\text{T}}\,\text{[GeV]}$ \\ \midrule
\texttt{Inclusive} & $\geq 6$ & $\geq 2$ & $\geq 500$\\ \hline
\bottomrule
\end{tabular}
\caption{
Summary of signal region criteria for ATLAS-CONF-2020-013, reproduced in full from Section 3 of \cite{ATLAS:20204t}.}
\end{table}
Jets are reconstructed using the anti-$k_t$ algorithm with radius parameter $R = 0.4$ \cite{Cacciari_2008}, and baseline leptons and jets are required to survive an overlap-removal procedure before confronting the selection criteria.

We have written code in C++ that can be run in the reconstruction (\texttt{-R}) mode of \textsc{MadAnalysis\,5} \cite{Conte_2014,Araz_2020} to emulate the analysis described above and allow us to apply it to new event samples. In all cases --- either for validation or for the analysis presented in \hyperref[s4]{Section 4} --- we provide as input some sample of hard-scattering events that have been matched to parton showers with the aid of \textsc{Pythia\,8} version 8.244 \cite{Pythia}, which also simulates hadronization. These showered and hadronized events are first passed by \textsc{MadAnalysis\,5} to \textsc{Delphes\,3} version 3.4.2 \cite{Delphes_OG,Delphes_3} and \textsc{FastJet} version 3.3.3 \cite{FJ}, which respectively model the response of the ATLAS detector and perform object reconstruction. For this step of the reimplementation, we use the \textsc{Delphes\,3} card shipped with \textsc{MadAnalysis\,5} for the ATLAS detector with anti-$k_t$ jet radius parameter appropriately set to $R = 0.4$. The reconstructed events are then analyzed by our reimplementation code, after which \textsc{MadAnalysis\,5} computes the acceptance of the event sample by the emulated selection criteria. With the acceptance(s) in hand, \textsc{MadAnalysis\,5} can use the CLs prescription \cite{Read:2002cls} to compute the upper limit at 95\% confidence level (CL) on the number of signal events given the official numbers of expected background events and observed events. 

The best practice is to validate a reimplemented analysis in some fashion before applying it to a new signal model. While this particular ATLAS analysis does not include e.g. detailed cut-flow charts, we can attempt a rough validation by simulating some Standard Model events included in the official event yields in Table 3 of \cite{ATLAS:20204t}. In particular, we simulate Standard Model production of four top quarks, which is the signal event for this analysis, and the three largest backgrounds reported by ATLAS to survive the signal region selection cuts. These are Standard Model top quark pair production in association with a $W$ boson ($t\bar{t}W$), with a $Z$ boson ($t\bar{t}Z$), and with a Higgs boson ($t\bar{t}H$). We simulate these events, and all other events for this work, in version 3.1.0 of \textsc{MadGraph5\texttt{\textunderscore}aMC@NLO} (\textsc{MG5\texttt{\textunderscore}aMC}) \cite{MG5,MG5_EW_NLO}, with showering and hadronization performed within \textsc{MG5\texttt{\textunderscore}aMC} by \textsc{Pythia\,8} as mentioned above. These events are simulated using the Universal FeynRules Output (UFO) implementation of the Standard Model shipped with \textsc{MG5\texttt{\textunderscore}aMC} \cite{UFO}. Our samples each include $5 \times 10^4$ ($2.5 \times 10^5$) $t\bar{t}W$, $t\bar{t}Z$ ($t\bar{t}H$, $t\bar{t}t\bar{t}$) events\footnote{We find that larger samples are required to achieve good statistical control for the latter two signatures.} with hard-scattering amplitudes at leading order (LO) in the strong coupling convolved with the NNPDF\,2.3 LO set of parton distribution functions \cite{nnpdf}. The top quarks in the $t\bar{t}t\bar{t}$ sample are decayed with the aid of \textsc{MadSpin}, a plugin for \textsc{MG5\texttt{\textunderscore}aMC}. The LO $t\bar{t}X$ ($X = W^{\pm},Z,H$) amplitudes are computed including the emission of up to two additional partons, in accordance with the procedure described by ATLAS. We convolve these hard-scattering amplitudes with the NNPDF\,2.3 set of parton distribution functions \cite{nnpdf} and then pass the showered and hadronized events to \textsc{MadAnalysis\,5} to begin the analysis procedure described above. The $t\bar{t}W$ sample is normalized to the SM cross section of $601\, \text{fb}$ computed at NLO in QCD that is used by ATLAS \cite{Campbell_2012}. The $t\bar{t}Z$ sample is combined with $t\bar{t}\gamma^*$ events and is normalized to the SM QCD NLO inclusive $t\bar{t} \ell^+ \ell^-$ cross section of $880\, \text{fb}$ \cite{LHCHiggsCrossSectionWorkingGroup:2016ypw}. The $t\bar{t}H$ sample is normalized to a cross section of $507.1\, \text{fb}$, which is the SM prediction at NLO in QCD with leading NLO electroweak corrections for a Higgs boson of $m_H = 125.0\, \text{GeV}$ \cite{LHCHiggsCrossSectionWorkingGroup:2016ypw}.

A comparison of our results and the official post-fit yields is available in \hyperref[t4]{Table 4}.
\begin{table}\label{t4}
\renewcommand\arraystretch{1.4}
\centering
\begin{tabular}{l|ccc}
\toprule
 & ATLAS yield  & \textsc{MadAnalysis\,5} yield & Error\,[\%] \\ \midrule
$t\bar{t}W + \text{jets}$ & $102\pm 26$ & $90.3$ & $-11.5$\\ \hline
$t\bar{t}Z + \text{jets}$ & $48\pm 9$ & $37.7$ & $-21.5$ \\ \hline
$t\bar{t}H + \text{jets}$ & $38 \pm 9$ & $40.2$ & $+5.73$ \\ \hline\hline
$t\bar{t}t\bar{t}$ [SM] & $30\pm 8$ & $28.4$ & $-5.48$ \\ \hline
\bottomrule
\end{tabular}
\caption{
Comparison of signal and leading background event yields reported in Table 3 of ATLAS-CONF-2020-013 \cite{ATLAS:20204t} and obtained using our reimplementation of this analysis in \textsc{MadAnalysis\,5}. $t\bar{t}t\bar{t}$ yields are normalized to SM cross section of $\sigma(pp \to t\bar{t}t\bar{t}) \approx 12\, \text{fb}$.}
\end{table}
We achieve errors of $\mathcal{O}(10)\%$, with the $t\bar{t}Z + \text{jets}$ background process suffering from the largest error. The acceptance of our simulated $t\bar{t}t\bar{t}$ events by our reimplemented analysis is roughly $1.65\%$, compared to an official acceptance of around $1.7\%$. The selection criterion with the lowest acceptance is the final-state lepton multiplicity cut listed in \hyperref[t2]{Table 2}, so we suspect that our lower efficiency in most channels is related to our treatment of final-state leptons somewhere in the simulation chain. Since our reimplementation is apparently somewhat less efficient than the official selection strategy, we expect the constraints it imposes on our color-octet scalar models to be marginally weaker than what could be obtained from a dedicated analysis by the ATLAS collaboration. Nevertheless we note here, before moving on, that the sgluon signal events we analyze in \hyperref[s4]{Section 4} are $0.6$--$1.3$\% efficient under this reimplementation, with scalar and pseudoscalar samples differing only statistically in this regard. The full results are displayed in \hyperref[f7]{Figure 7}. Finally, as another means of validation, we include the jet and $b$-tagged jet multiplicity distributions for our same SM $t\bar{t}t\bar{t}$ sample in \hyperref[f8]{Figure 8}, which compares these distributions between ATLAS' data, ATLAS' pre- and post-fit SM $t\bar{t}t\bar{t}$ predictions, and (viz. \hyperref[s4]{Section 4}) a sample corresponding to a color-octet scalar $t\bar{t}t\bar{t}$ contribution at a benchmark point apparently well suited to fit the reported excess. In this figure we demonstrate good agreement between our sample, normalized to the event yield reported in \hyperref[t4]{Table 4}, and the ATLAS $t\bar{t}t\bar{t}$ results normalized to the SM (pre-fit) cross section.

\subsection{ATLAS-CONF-2020-002: search for new phenomena with many jets + $E^{\mathrm{miss}}_{\mathrm{T}}$}
\label{s3.2}

This second analysis, which was announced in ATLAS-CONF-2020-002, looks for new phenomena in final states with large numbers of jets and significant missing transverse energy. It particularly targets events with at least eight anti-$k_t$ radius $R=0.4$ jets with transverse momentum $p_{\text{T}} > 50\, \text{GeV}$ or higher, depending on signal region. It also requires a high missing transverse energy significance $\mathcal{S}(E_{\text{T}}^{\text{miss}})$ in order to disambiguate genuine $E_{\text{T}}^{\text{miss}}$ associated with non-interacting particles from specious missing energy due to mismeasurements and fluctuations. This search vetoes virtually all leptons surviving an overlap-removal procedure. The final noteworthy element of this search is a set of cuts on the cumulative mass $M^{\Sigma}_{\text{J}}$ of high-$p_{\text{T}}$ large-radius (anti-$k_t$ radius $R=1.0$) jets, which is intended to stringently control the SM multijet background. ATLAS performed a multi-bin and a single-bin subanalysis, the latter of which defines eight non-overlapping signal regions and which we reinterpret using the \textsc{MadAnalysis\,5} framework as described for the previous analysis. We summarize the preselection criteria, the common selections, and the signal region criteria for the single-bin subanalysis in Tables \hyperref[t5]{5}--\hyperref[t7]{7}.

\begin{table}\label{t5}
\renewcommand\arraystretch{1.4}
\centering
\begin{tabular}{c|c c c c c}
\toprule
Criterion & Electrons  & Muons & Photons & Jets & $b$-tagged jets\\
\midrule
$p_{\text{T}}\,\text{[GeV]}$ & $> 7.0$ & $> 6.0$ & $> 40$ & \makecell{$>20\ \text{[$R = 0.4$]}$\\ $>100\ \text{[$R = 1.0$]}$} &  $>20$\\\hline
$|\eta$| & $<2.47$ & $< 2.7$ & \makecell{$< 2.37$\\ $\text{and} \notin (1.37,1.52)$} & \makecell{$<2.8\ \text{[$R = 0.4$]}$\\ $<2.0\ \text{[$R = 1.0$]}$} & $<2.5$ \\\hline
\bottomrule
\end{tabular}
\caption{
Summary of preselection criteria in ATLAS-CONF-2020-002, reproduced in part from Section 4 of \cite{ATLAS:2020new}.}
\end{table}

\begin{table}\label{t6}
\renewcommand\arraystretch{1.4}
\centering
\begin{tabular}{c|c c}
\toprule
Selection criterion & \multicolumn{2}{c}{Selection ranges} \\
\midrule
Jet multiplicity, $N_{\text{jet}}$
& $N_{\text{jet}}^{50} \geq \big\{8, 9, 10, 11, 12\big\}$  & $N_{\text{jet}}^{80} \geq 9$ \tabularnewline \hline
Trigger thresholds & 6 or 7 jets, $E_{\text{T}} > 45\,\text{GeV}$ & 5 jets, $E_{\text{T}} > 65\ \text{or}\ 70\,\text{GeV}$  \\ \hline
Lepton veto 			        & \multicolumn{2}{c}{0 baseline leptons, $p_{\text{T}}>10\,\text{GeV}$} \\\hline
$E_{\text{T}}^{\text{miss}}$ significance, $\mathcal{S}(E_{\text{T}}^{\text{miss}})$
& \multicolumn{2}{c}{$\mathcal{S}(E_{\text{T}}^{\text{miss}})> 5.0$} \\\hline
\bottomrule
\end{tabular}
\caption{
Summary of common selection criteria in ATLAS-CONF-2020-002, reproduced in part from Sections 4 and 5 and Table 1 of \cite{ATLAS:2020new}. Variable trigger thresholds depend on year in which data were collected.}
\end{table}

\begin{table}\label{t7}
\renewcommand\arraystretch{1.4}
\centering
\begin{tabular}{l|cccc}
\toprule
Signal region & $N_{\text{jet}}^{50}$ & $N_{\text{jet}}^{80}$  & $N_{b\text{-jet}}$ & $M_{\text{J}}^{\Sigma}\,\text{[GeV]}$ \\ \midrule
\texttt{SR-8ij50-0ib-MJ500} & ~~$\geq 8$ & - & - & $\geq 500$ \\
\texttt{SR-9ij50-0ib-MJ340} & ~~$\geq 9$ & - & - & $\geq 340$ \\
\texttt{SR-10ij50-0ib-MJ340} & $\geq 10$ & - & - & $\geq 340$ \\
\texttt{SR-10ij50-0ib-MJ500} & $\geq 10$ & - & - & $\geq 500$ \\
\texttt{SR-10ij50-1ib-MJ500} & $\geq 10$ & - & $\geq 1$ & $\geq 500$ \\
\texttt{SR-11ij50} & $\geq 11$ & - & - & - \\
\texttt{SR-12ij50-2ib} & $\geq 12$ & - & $\geq 2$ & - \\
\texttt{SR-9ij80} & - & $\geq 9$ & - & - \\ \hline
\bottomrule
\end{tabular}
\caption{
Summary of signal region criteria for single-bin selections in ATLAS-CONF-2020-002, reproduced in full from Table 3 of \cite{ATLAS:2020new}. A dash (-) indicates that no requirement is applied to the corresponding variable.
The requirement $\mathcal{S}(E_{\text{T}}^{\text{miss}})>5.0$ is applied to all bins.}
\end{table}
\renewcommand\arraystretch{1.0}

In addition to providing model-independent limits at $95\%$ CL on the number of bSM events in each signal region of the single-bin analysis, ATLAS used the multi-bin analysis to constrain several benchmark bSM models. One of these benchmarks is a model of gluino-mediated top quark production in which pair-produced gluinos decay with unit branching fraction to $t\bar{t}+ \tilde{\chi}^0_1$ (with $\tilde{\chi}^0_1$ a suggestively labeled neutralino) via a highly off-shell squark $\tilde{q}$, generating a four-top signature from the process $pp \to \tilde{g}\tilde{g} \rightarrow t\bar{t}t\bar{t} + E_{\text{T}}^{\text{miss}}$. The multi-bin analysis, which ATLAS says provides the most stringent constraints with few exceptions, disfavors gluinos with $m_{\tilde{g}} \lesssim 1.8\, \text{TeV}$ in scenarios with light $\tilde{\chi}^0_1$, with limits relaxing as $m_{\tilde{\chi}^0_1}$ grows (viz. Figure 10(b) of the analysis). In order to validate our reimplementation of the single-bin analysis, we attempt to reproduce ATLAS' constraints on this benchmark model, keeping in mind that an accurate emulation of the single-bin analysis should underexclude relative to the official multi-bin results.

There is a simplified model of supersymmetric quantum chromodynamics \cite{Degrande_2016} that was implemented in the \textsc{Mathematica}$^\copyright$\ package \textsc{FeynRules} \cite{Mathematica,FR_OG,FR_2}, whose UFO output is publicly available on the \textsc{FeynRules} model database, that is well suited to simulate the signal events needed to validate our reimplementation. We have straightforwardly modified this UFO to provide the desired three-body gluino decay $\tilde{g} \to t\bar{t} + \tilde{\chi}^0_1$ with unit branching fraction. The stop squarks that mediate this decay are decoupled at $m_{\tilde{t}_{\text{L}}} = 9.0\, \text{TeV}$ and $m_{\tilde{t}_{\text{R}}} = 8.0\, \text{TeV}$. We simulate $10^4$ gluino pair-production and decay events for a variety of gluino and neutralino masses in \textsc{MG5\texttt{\textunderscore}aMC} in much the same way as we produce the signal events for ATLAS-CONF-2020-013. The samples are normalized to the cross sections of gluino pair production at approximate next-to-next-to-leading order (NNLO) in QCD, including soft gluon emission resummation at next-to-next-to-leading-logarithmic (NNLL) accuracy, that are used by ATLAS \cite{Beenakker_2016,Beenakker_2014,Beenakker_2009}. For this reimplementation, we use a \textsc{Delphes\,3} card for the ATLAS detector modified to include a collection of jets for both anti-$k_t$ radius parameters ($R = 0.4$ and $R=1.0$) required for this search.

\begin{figure}\label{f4}
\centering
\includegraphics[scale=0.7]{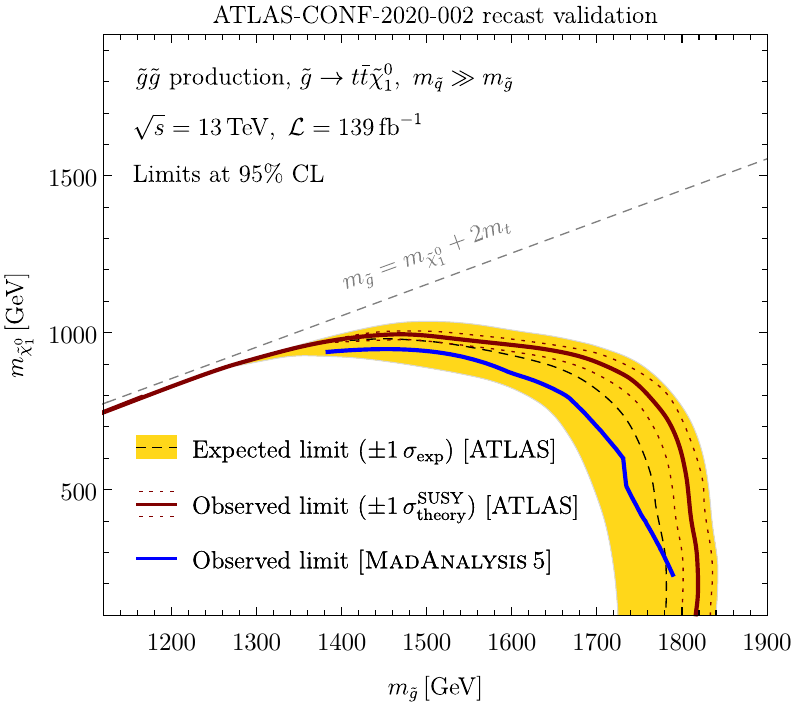}
\caption{Exclusion limits on gluino pair production in the $(m_{\tilde{g}},m_{\tilde{\chi}^0_1})$ plane, assuming gluino decays to $t\bar{t} + \tilde{\chi}^0_1$ via a virtual squark $\tilde{q}$. Red and black/yellow results were obtained from Figure 10(b) of ATLAS-CONF-2020-002 \cite{ATLAS:2020new}. The blue exclusion curve is imposed on the benchmark model described in the text by our reimplementation of this analysis, and constitutes our validation of this reimplementation. Our limits are weaker than those of the official analysis, but this is to be expected --- see text. At any rate, discrepancies are of $\mathcal{O}(10)\%$.}
\end{figure}

A comparison of our results and the official benchmark model limits at $95\%$ CL is available in \hyperref[f4]{Figure 4}. We find that the most sensitive signal region in almost all cases is \texttt{SR-10ij50-1ib-MJ500}, which is indeed one of the two signal regions optimized for this signal according to ATLAS. The acceptances of our simulated gluino pair-production events by this signal region vary from $0.3\%$ to $3\%$ with increasing $m_{\tilde{g}}$. For this reimplementation we achieve errors of $\mathcal{O}(10)\, \text{GeV}$ (up to around $100\, \text{GeV}$) in our observed limits at $95\%$ CL. The largest errors by this metric amount to $-5\%$. It is unclear how much of this error can be ascribed to the inferior sensitivity of the single-bin subanalysis and how much is genuine error in our reimplementation. Another possible source of error is our implementation of the cut on $E_{\text{T}}^{\text{miss}}$ significance. ATLAS has begun to use an ``object-based'' definition \cite{ATLAS:SMET} of $\mathcal{S}(E_{\text{T}}^{\text{miss}})$ (in which this quantity is dimensionless) that to our knowledge cannot be implemented in \textsc{MadAnalysis\,5} at this time. In keeping with some validated reimplemented searches available on the \textsc{MadAnalysis\,5} Public Analysis Database (PAD) \cite{Dumont_2015} that have confronted this same problem, we have used a $\mathcal{S}(E_{\text{T}}^{\text{miss}})$ proxy,
\begin{align}\label{e3.2}
\mathcal{S}_{\text{proxy}}(E_{\text{T}}^{\text{miss}}) = \frac{E_{\text{T}}^{\text{miss}}}{\sqrt{H_{\text{T}}}}\ \ \ \text{with}\ \ \ H_{\text{T}} = \sum_i p_{\text{T}i}^{\text{jet}},
\end{align}
which was used by ATLAS prior to the adoption of the new object-based definition \cite{Aaboud_2017}. (This proxy has units of $\text{GeV}^{1/2}$, so our cut is at $\mathcal{S}_{\text{proxy}}(E_{\text{T}}^{\text{miss}}) = 5\, \text{GeV}^{1/2}$.) Nevertheless, we consider the agreement good enough to proceed. This analysis reimplementation is considered validated and is available for public download, along with a somewhat more detailed validation note, on the aforementioned Public Analysis Database \cite{DVN/0UHTPC_2021}. We conclude by reporting that the sgluon signal events we analyze in \hyperref[s4]{Section 4} are up to about 1\% efficient in \texttt{SR-10ij50-1ib-MJ500} and 6\% efficient in \texttt{SR-8ij-0ib-MJ500} under this reimplementation, with scalar and pseudoscalar samples again showing negligible discrepancies. The full results are displayed in \hyperref[f7]{Figure 7}.
\section{Fitting the $\boldsymbol{t\bar{t}t\bar{t}}$ excess today and in the future}
\label{s4}

In this section, as advertised, we examine the contributions of either color-octet scalar to the observed cross section $\sigma(pp \to t \bar{t} t \bar{t})$ of four-top quark production at the LHC. A schematic diagram for these processes is displayed in \hyperref[f5]{Figure 5}.
\begin{figure}\label{f5}
\begin{align*}
\scalebox{0.75}{\begin{tikzpicture}[baseline={([yshift=-.5ex]current bounding box.center)},xshift=12cm]
\begin{feynman}[large]
\vertex (i1);
\vertex [right = 1.25cm of i1, blob] (i2){};
\vertex [below left = 0.15cm of i2] (q1);
\vertex [above left=0.46 cm of q1] (r1);
\vertex [below left =0.15cm of q1] (q2);
\vertex [above left = 0.39cm of q2] (r2);
\vertex [above left=0.15cm of i2] (l1);
\vertex [below left = 0.46cm of l1] (r3);
\vertex [above left=0.15cm of l1] (l2);
\vertex [below left=0.39cm of l2] (r4);
\vertex [above=0.2cm of i2] (q3);
\vertex [above=0.2cm of q3] (q4);
\vertex [above left=1.75cm of i2] (v3);
\vertex [below left=0.15cm of v3] (p1);
\vertex [below left =0.15cm of p1] (p2);
\vertex [below left=1.75cm of i2] (v4);
\vertex [above left=0.15cm of v4] (p3);
\vertex [above left =0.15cm of p3] (p4);
\vertex [above right=0.9 cm and 2.4cm of i2,blob] (v1){};
\vertex [above right=0.6 cm and 2cm of v1] (t1);
\vertex [below right=0.6cm and 2cm of v1] (t2);
\vertex [below right=0.9cm and 2.4cm of i2,blob] (v2){};
\vertex [above right=0.6 cm and 2cm of v2] (t3);
\vertex [below right=0.6 cm and 2cm of v2] (t4);
\diagram* {
(p1) -- [ultra thick] (r1),
(p2) -- [ultra thick] (r2),
(p3) -- [ultra thick] (r3),
(p4) -- [ultra thick] (r4),
(v3) -- [ultra thick] (i2),
(v4) -- [ultra thick] (i2),
(i2) -- [ultra thick, scalar ] (v1),
(i2) -- [ultra thick, scalar] (v2),
(t2) -- [ultra thick, fermion] (v1) -- [ultra thick, fermion] (t1),
(t3) -- [ultra thick, fermion] (v2) -- [ultra thick, fermion] (t4),
};
\end{feynman}
\node at (0.25,1.35) {$p$};
\node at (0.25,-1.4) {$p$};
\node at (2.8,0.82) {$O$};
\node at (2.8,-0.82) {$O$};
\node at (6.4,1.55) {$t$};
\node at (6.4,0.3) {$\bar{t}$};
\node at (6.4,-0.3) {$\bar{t}$};
\node at (6.4,-1.55) {$t$};
\end{tikzpicture}}
\end{align*}
\caption{Schematic diagram for the contribution of the scalar sgluon to the four-top cross section $\sigma(pp \to t\bar{t}t\bar{t})$ at the LHC. Diagrams for effective vertices are displayed in Figures \hyperref[f1]{1} and \hyperref[f2]{2}.}
\end{figure}
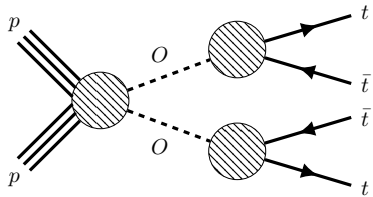The maximum-likelihood fit performed by ATLAS in ATLAS-CONF-2020-013 gives a signal strength of \cite{ATLAS:20204t}
\begin{align}\label{e4.1}
 \mu \equiv \frac{\sigma(pp \to t\bar{t}t\bar{t})}{\sigma_{\text{SM}}(pp \to t\bar{t}t\bar{t})} = 2.0_{-0.6}^{+0.8}
\end{align}
and an observed post-fit yield of 60 signal events (viz. \hyperref[t4]{Table 4}). Taken together, these results imply an excess over the Standard Model of $30^{+24}_{-18}$ events, about half the post-fit yield, and $\mathcal{O}(10)\,\text{fb}$. The ATLAS jets + $E_{\text{T}}^{\text{miss}}$ search, on the other hand, finds no excess over Standard Model expectation. In this analysis, we look at which regions of color-octet scalar parameter space are ruled out by these measurements and which regions can supply an excess of four-top events, and we investigate whether sgluons capable of fitting the excess are currently ruled out or could be discovered or excluded in the future. The sgluon-mediated contributions to $\sigma(pp \to t\bar{t}t\bar{t})$ are approximately the product of the sgluon pair-production cross sections and their branching fractions to $t\bar{t}$, since both particles easily satisfy the narrow-width approximation below the on-shell squark thresholds \cite{Carpenter:2020mrsm,Carpenter2021coloroctet}. It is therefore straightforward to derive constraints on color-octet scalars in both the fundamental parameter space of minimal Dirac gaugino models, in which these branching fractions have been calculated, and in generic parameter space described by the sgluons' branching fractions to $t\bar{t}$.

In order to accurately compute the pair-production cross sections and to generate simulated events for analysis in the \textsc{MadAnalysis\,5} framework discussed in \hyperref[s3]{Section 3}, we have implemented the effective model defined in \hyperref[s2.2]{Section 2.2} in \textsc{FeynRules} version 2.3.43 within \textsc{Mathematica}$^\copyright$\ version 12.0 \cite{FR_OG,FR_2,Mathematica}. We have formulated a bare Lagrangian and used \textsc{FeynRules} to initiate QCD renormalization at NLO by defining all counterterms of $\mathcal{O}(g_3^4) = \mathcal{O}(\alpha_3^2)$. We have then used \textsc{NloCT} version 1.02 \cite{NLOCT}, which is shipped with \textsc{FeynRules}, to compute all one-loop QCD counterterms pursuant to a set\footnote{We have retained the default renormalization conditions: physical fields and masses are renormalized in the on-shell scheme, the strong coupling is renormalized in the zero-momentum scheme, and other parameters are renormalized in $\overline{\text{MS}}$.} of renormalization conditions. All necessary one-loop amplitudes have been evaluated using the \textsc{FeynRules} interface to version 3.11 of the diagram-generating package \textsc{FeynArts} \cite{FA}. With these counterterms in hand, we have finally used \textsc{FeynRules} once more to generate a Universal FeynRules Output (UFO) \cite{UFO} capable of Monte Carlo event simulation at NLO in the strong coupling. We have used this UFO as input for the same \textsc{MG5\texttt{\textunderscore}aMC} + \textsc{Pythia\,8} setup used to validate our recasts as described in \hyperref[s3]{Section 3}.

\begin{figure}\label{f6}
\centering
\includegraphics[scale=1]{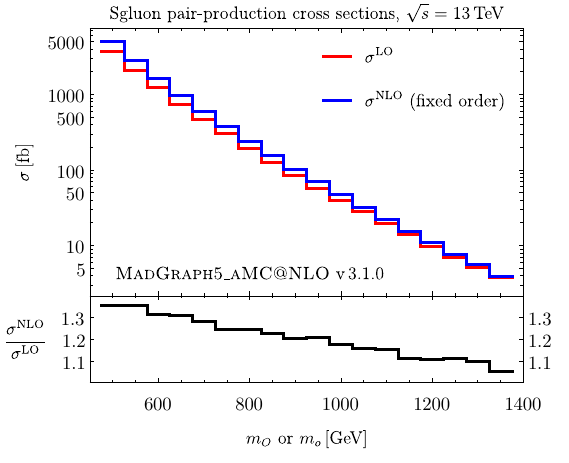}
\caption{Cross sections of pair production of scalar or pseudoscalar sgluons. In the upper panel we display results at leading order (LO) and next-to-leading order (NLO) in the strong coupling. The lower panel shows the $K$ factor for this process, given by the ratio of the NLO result to the LO result.}
\end{figure}

We first use this machinery to compute the total cross sections of pair production for either sgluon in a broad range of sgluon masses. (Recall that their pair-production cross sections are identical in these models.) The results are displayed in \hyperref[f6]{Figure 6}.
This figure shows results at LO and NLO in the strong coupling and the NLO enhancement ($K$) factor given by the ratio of the latter to the former for each chosen sgluon mass. The results at both orders are obtained by convolving the scattering amplitudes with the NNPDF\,2.3 set of parton distribution functions \cite{nnpdf}, with renormalization and factorization scales fixed at the mass of the pair-produced sgluon. Our results are consistent with those of \cite{Degrande:2015pprod}, which were also obtained using the $\textsc{FeynRules} \to \textsc{MG5\texttt{\textunderscore}aMC}$ chain; and with those of \cite{Carpenter:2020mrsm}, which were computed in \textsc{Mathematica}$^{\copyright}$\ version 12.0. We see that the cross section plummets from $\mathcal{O}(10^3)\, \text{fb}$ to $\mathcal{O}(1)\, \text{fb}$ in the displayed sgluon mass range, while the $K$ factor diminishes gently from $\sim\! 1.33$ to $\sim\! 1.05$. These results give us considerable parameter space in which to fit an $\mathcal{O}(10)\, \text{fb}$ excess.

\begin{figure}\label{f7}
\centering
\includegraphics[scale=0.75]{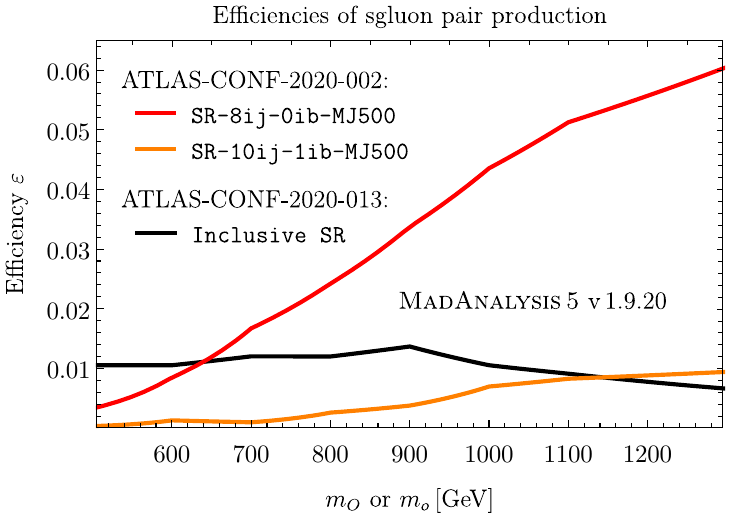}
\caption{Efficiencies of event samples of scalar or pseudoscalar sgluon pair production followed by decays to $t\bar{t}$ pairs under relevant signal regions in ATLAS-CONF-2020-002 and ATLAS-CONF-2020-013. Efficiencies hover around 1\% in the inclusive SR for ATLAS-CONF-2020-013, the measurement of $\sigma(pp \to t\bar{t}t\bar{t})$, and rise with increasing sgluon mass for both displayed SRs of ATLAS-CONF-2020-002, the search for new phenomena.}
\end{figure}

We then prepare event samples for analysis by generating the amplitudes for sgluon pair production followed by decays to $t\bar{t}$ at LO in QCD, convolving these with the same set of parton distribution functions at the same scales used for the cross section calculations, and matching these hard-scattering events with parton showers with the aid of \textsc{Pythia\,8}, which also simulates hadronization. We simulate $10^4$ events for each benchmark point, generally chosen at intervals of $m_O$ or $m_o$ of $100\, \text{GeV}$ except where finer detail is desired. We then use these showered event samples as inputs for the reconstruction mode of \textsc{MadAnalysis\,5}. As described in \hyperref[s3]{Section 3}, the response of the ATLAS detector is modeled by \textsc{Delphes\,3} with object reconstruction performed by \textsc{FastJet}, and the appropriate selection criteria are subsequently imposed by our reimplementations of ATLAS-CONF-2020-002 and ATLAS-CONF-2020-013. \textsc{MadAnalysis\,5} finally computes the upper limits at $95\%$ CL on the allowed number of events from our signal, given the numbers of expected background events and total observed events, for all signal regions in both analyses. We previewed the efficiencies of these searches in \hyperref[s3]{Section 3}, but in \hyperref[f7]{Figure 7} we provide more specific results, showing the acceptances of our event samples by the inclusive signal region of ATLAS-CONF-2020-013 and the two signal regions of ATLAS-CONF-2020-002 that impose the most stringent expected or observed limits at 95\% CL.
We present these efficiencies concurrently for the scalar and pseudoscalar sgluons since, as we noted earlier, we find negligible discrepancies between scalar and pseudoscalar results. We obtain efficiencies of 0.6--1.3\% for the former analysis and varying but generally higher efficiencies --- up to $6\%$ --- for the latter search. We note that the latter efficiencies rise with increasing sgluon mass: this is because heavier sgluons decay to increasingly boosted top quarks, so these events are better able to survive the stringent cuts on jet $p_{\text{T}}$ and $E_{\text{T}}^{\text{miss}}$ designed to control the Standard Model background. We show below that this has tangible effects on the relative sensitivities of the two ATLAS analyses.

\subsection{Results from LHC Run 2}
\label{s4.1}

\begin{figure}\label{f8}
\centering
\includegraphics[scale=0.7]{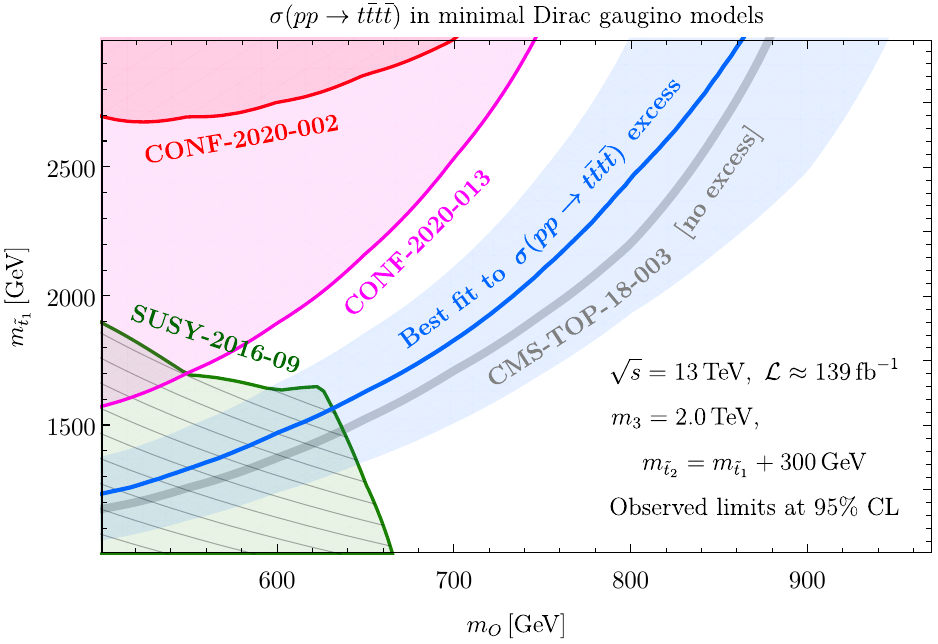}
\caption{Map of scalar sgluon parameter space, including regions excluded by relevant analyses and favored by the excess in $\sigma(pp \to t\bar{t}t\bar{t})$ reported by ATLAS. Here the scalar sgluon and stop squark masses are varied while the Dirac gluino and stop mass splitting are fixed. The scalar branching fraction to top quarks $\text{BF}(O \to t\bar{t})$ increases with the ordinate. For context, also displayed are limits from the most recent CMS measurement, which sees no excess. See \hyperref[f9]{Figure 9} and subsequent plots for expanded maps of color-octet scalar parameter space covering a wider range of scenarios.}
\end{figure}

\begin{figure}\label{f9}
\centering
\includegraphics[scale=0.7]{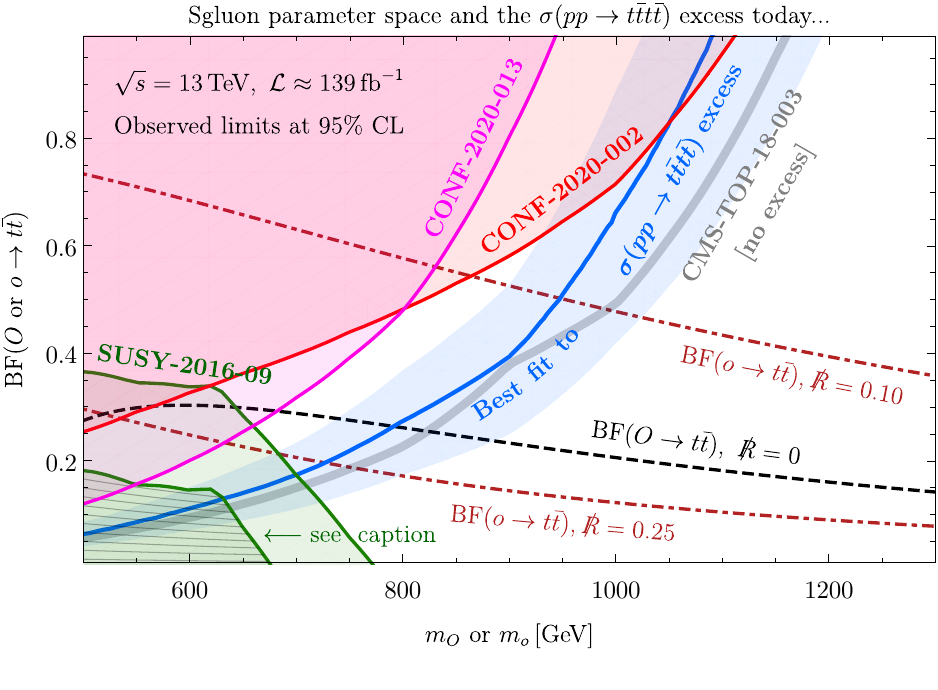}
\caption{Expanded map of color-octet scalar parameter space, including regions excluded by relevant analyses and favored by the excess in $\sigma(pp \to t\bar{t}t\bar{t})$ reported by ATLAS. This and subsequent plots are in a generic plane of color-octet scalar mass and branching fraction to top quarks. Dashed lines trace branching fractions to $t\bar{t}$ for indicated sgluon in specific scenarios described in text. Larger green region applies to pseudoscalar, which has no decays involving an electroweak gauge boson. Smaller hatched green region (visible also in \hyperref[f8]{Figure 8}) applies to scalar, which does enjoy such decays, thus weakening its branching fraction to gluons. Limits from most recent CMS measurement finding no excess are again displayed for context. See \hyperref[f10]{Figure 10} for HL-LHC projections in the same parameter space.}
\end{figure}

We finally turn to fits and constraints in the parameter space of our models in light of both four-top final-state event analyses in Figures \hyperref[f8]{8} and \hyperref[f9]{9}. The first of these shows a plot in the plane of scalar sgluon mass $m_O$ and lightest stop squark mass $m_{\tilde{t}_1}$.
This is a fairly realistic region in the natural parameter space of minimal ($R$-symmetric) models with Dirac gauginos in which the Dirac gluino mass is fixed at $m_3 = 2.0\, \text{TeV}$ and the heavier stop squark $\tilde{t}_2$ is $300\,\text{GeV}$ more massive than $\tilde{t}_1$. As we noted in \hyperref[s2.2]{Section 2.2}, and showed in \hyperref[f3]{Figure 3}, the branching fraction $\text{BF}(O \to t\bar{t})$ of the scalar sgluon to top quarks increases with increasing $m_{\tilde{t}_1}$ (and $m_{\tilde{t}_2}$) relative to fixed $m_3$. The top edge of this plot corresponds to branching fractions of around thirty percent. The regions disfavored in this parameter space by the two analyses we have reinterpreted are rendered in red and magenta\footnote{We have alluded to this above, but the signal regions of ATLAS-CONF-2020-002 that impose these limits are \texttt{SR-8ij-0ib-MJ500} and, more often, \texttt{SR-10ij-1ib-MJ500}.}. The green region in the lower left-hand corner shows parameter space excluded by ATLAS-SUSY-2016-09, an ATLAS search using $36.7\, \text{fb}^{-1}$ of $pp$ collisions at $\sqrt{s} = 13\, \text{TeV}$ for pair-produced resonances in flavorless four-jet final states \cite{ATLAS:2018s1} that was used to constrain pair-produced color-octet scalars decaying to gluons. This search naturally becomes more sensitive than either search involving final states with top quarks for regions of parameter space where $\text{BF}(O \to t\bar{t})$ is too low. Finally visible in \hyperref[f8]{Figure 8} is a blue line with associated error band tracing the combinations of sgluon masses and squark and gluino spectra capable of fitting the $\sigma(pp \to t\bar{t}t\bar{t})$ excess of roughly thirty events\footnote{The line gives exactly thirty events, corresponding to the mean signal strength of $\mu = 2.0$. The left side of the band accommodates excesses of up to 54 events; the right side allows as few as 12 additional events.} announced by ATLAS (viz. \eqref{e4.1} and surrounding discussion). This line corresponds to cross sections of 17--$22\, \text{fb}$ depending on the efficiencies of the events under ATLAS-CONF-2020-013 (viz. \hyperref[f7]{Figure 7}). We see that only the search for four flavorless jets currently excludes any of the displayed parameter space well suited to fit the measured excess, leaving scalars of mass $m_O \in (650,850)\,\text{GeV}$ as viable candidates. In the interest of completeness, we note that --- despite the appearance of \hyperref[f8]{Figure 8} --- there are some regions left open by these searches in low-mass scalar sgluon parameter space, so this figure should not be understood to imply that such particles are universally excluded below $m_O \approx 600$--$700\, \text{GeV}$. In particular, ATLAS-SUSY-2016-09 only applies to resonances heavier than $500\, \text{GeV}$, which is just out of frame of \hyperref[f8]{Figure 8} \cite{ATLAS:2018s1}. The searches for particles decaying to $t\bar{t}$, on the other hand, are ineffective below the on-shell top quark pair decay threshold at about $350\, \text{GeV}$. A slightly outdated but more comprehensive survey of collider constraints on sgluons in minimal $R$-symmetric models is available in \cite{Carpenter:2020mrsm}.

We finally note the gray band running through \hyperref[f8]{Figure 8}, which traces the exclusion limit at 95\% CL from the most recent search for four-top quark production by the CMS collaboration, which we mentioned in the Introduction and was announced in CMS-TOP-18-003 \cite{CMS_4t_recent}. This analysis was performed on $137\,\text{fb}^{-1}$ of $pp$ collisions, constituting almost the full Run 2 dataset; it targets multilepton final states fairly similar to those considered in ATLAS-CONF-2020-013. As we alluded to in the Introduction, however, CMS-TOP-18-003 reports no significant excess, instead measuring a cross section of $\sigma(pp \to t\bar{t}t\bar{t}) = 12.6^{+5.8}_{-5.2}\,\text{fb}$ with an observed significance relative to background of 2.6 standard deviations. We include the limits imposed by the CMS search in order to provide a complete view of the current experimental landscape that shows the tension between the current ATLAS and CMS results. These limits are computed in much the same way as the others in this and following figures using the \textsc{MadAnalysis}\,5 framework; in particular, there is already a validated implementation of CMS-TOP-18-003 on the Public Analysis Database, which we straightforwardly apply to our sgluon event samples \cite{DVN/OFAE1G_2020}. We find, generally consistent with our expectations, that the CMS analysis disfavors most of the parameter space well suited to fit the excess suggested by ATLAS-CONF-2020-013, including all space providing thirty or more events in excess of the SM prediction.

\hyperref[f9]{Figure 9}, by contrast with \hyperref[f8]{Figure 8}, shows a plot in the plane of sgluon masses and branching fraction to $t\bar{t}$. This parameter space is more generic and less tailored to the frequently considered scenario of CP-even sgluons in minimal $R$-symmetric models. It offers a broader view of color-octet scalar parameter space encompassing a variety of scenarios that apply to both scalar and pseudoscalar sgluons. The branching fractions at each point in this plot should be understood to be functions of stop squark and gluino masses, the extent of $R$ symmetry breaking, and the binary choice of sgluon CP. Since the scalar and pseudoscalar states have identical pair-production cross sections and closely similar acceptances of events by the two ATLAS analyses, we have in this plot provided statistically concurrent limits for both particles. There is one exception to this rule: the green region in the lower left-hand corner of \hyperref[f9]{Figure 9}, which again shows the results of ATLAS-SUSY-2016-09, must be interpreted differently for each species. The larger green region applies if $\text{BF}(O\, \text{or}\, o \to gg) + \text{BF}(O\,\text{or}\,o \to t\bar{t}) \approx 1$; i.e., if no other decays are non-negligible in this mass range. This is the case for e.g. a pseudoscalar sgluon in Dirac gaugino models with broken $R$ symmetry. The smaller hatched green region, by contrast, applies to minimal $R$-symmetric models wherein the scalar sgluon decays to a gluon and a photon or $Z$ boson at roughly $30\%$ of the rate of decay to gluon pairs. The presence of this decay channel diminishes the branching fraction $\Gamma(O \to gg)$ to gluons for a given $\Gamma(O \to t\bar{t})$, weakening the sensitivity of this search in this scenario. The hatched region in \hyperref[f9]{Figure 9} is in direct correspondence to the green region in \hyperref[f8]{Figure 8}, which applies solely to the scalar sgluon. In fact, since $\text{BF}(O \to t\bar{t})$ is controlled by the squark-gluino hierarchy, \hyperref[f8]{Figure 8} is effectively subsumed by roughly the lower left quadrant of \hyperref[f9]{Figure 9}. Recall, on the other hand, that the pseudoscalar sgluon in the $R$-symmetric limit has unit branching fraction to $t\bar{t}$, so \hyperref[f9]{Figure 9} extends \hyperref[f8]{Figure 8} to include constraints on pseudoscalars in minimal Dirac gaugino models.

The last important features of \hyperref[f9]{Figure 9} are the three dashed lines showing branching fractions to top quarks for one sgluon or the other in three specific scenarios. The black dashed line corresponds to a slice of \hyperref[f3]{Figure 3} at $m_{\tilde{t}_1} = 3.0\, \text{TeV}$ (hence $m_{\tilde{t}_2} = 3.3\, \text{TeV}$; recall also $m_3 = 2.5\, \text{TeV}$). This line comes fairly close to tracing the top edge of \hyperref[f8]{Figure 8}. This is an interesting region in the parameter space of minimal $R$-symmetric (Dirac gaugino) models where the usual squark-gluino hierarchy is reversed, with quite heavy stops, in order to produce $\text{BF}(O \to t\bar{t})$ compatible with the $\sigma(pp \to t\bar{t}t\bar{t})$ excess while avoiding constraints from ATLAS-SUSY-2016-09. The red dashed lines, meanwhile, show the pseudoscalar branching fractions to top quarks in the two benchmark scenarios with low-to-moderate $R$ symmetry breaking considered in \cite{Carpenter2021coloroctet}. In these scenarios --- as we discussed in \hyperref[s2.3]{Section 2.3} --- $R$ symmetry breaking allows the pseudoscalar to decay to gluon pairs, thus diminishing $\text{BF}(o \to t\bar{t})$ from its $R$-symmetric value of near unity. All three of these lines are merely interesting specific choices; they can all be smoothly adjusted in either direction by making different parameter choices.

Taken together, the ATLAS measurement of $\sigma(pp \to t\bar{t}t\bar{t})$ and the search for new phenomena with jets + $E_{\text{T}}^{\text{miss}}$ exclude regions with both low and high sgluon masses and branching fractions to $t\bar{t}$. A sizable region of parameter space with a broad range of $t\bar{t}$ branching fractions and mass $m_O \in (650,1000)\, \text{GeV}$ or $m_o \in(750,1000)\,\text{GeV}$ --- depending, recall, on each particle's branching fraction to $gg$ --- fits a signal strength of $\mu=2.0$ relative to the Standard Model prediction for $\sigma(pp \to t\bar{t}t\bar{t})$. All three benchmark scenarios represented by dashed lines in \hyperref[f9]{Figure 9} look promising; a pseudoscalar sgluon with $0.10 < \slashed{R} < 0.25$ is particularly intriguing, since it can attain the indicated branching fractions in spectra with light squarks \cite{Carpenter2021coloroctet}. The jets + $E_{\text{T}}^{\text{miss}}$ search, however, does impinge on the upper end of the parameter space well suited to fit the excess, excluding $\sim\!1\, \text{TeV}$ sgluons decaying to $t\bar{t}$ more than about $80\%$ of the time. This result notably excludes a pseudoscalar sgluon in minimal Dirac gaugino models --- which has $\text{BF}(o \to t\bar{t}) \approx 1$ --- below $m_o \approx 1105\, \text{GeV}$ (though this constraint is only valid above the threshold for decays to top quarks). We emphasize that pseudoscalar sgluons heavier than this limit do not fit the central value of the measured excess, but they can still fit in the error band. On the other hand, as we discussed in \hyperref[s3.2]{Section 3.2}, our reimplementation of ATLAS-CONF-2020-002 is less sensitive than the official multi-bin analysis, so the true limits from that search may rule out some more high-branching fraction parameter space. It is also worth noting that this limit is fairly close to that imposed by a previous CMS measurement \cite{CMS:20184t,Darme:2018rec,darme2021topphilic} of $\sigma(pp \to t\bar{t}t\bar{t})$ with a lower cross section and lower significance than the most recent ATLAS results. Finally, we observe that there is a point around $m_O$ or $m_o \approx 800\, \text{GeV}$ where ATLAS-CONF-2020-002 becomes more sensitive to our signals than the four-top analysis. This transition is a reflection of the efficiencies of the relevant signal regions of ATLAS-CONF-2020-002, which we showed in \hyperref[f7]{Figure 7} grow with increasing sgluon mass. Below this region, as we discussed above, ATLAS-CONF-2020-013 --- which looks for top quarks produced near threshold, as in the Standard Model --- becomes most sensitive. But we reiterate that despite the complementarity between these two analyses of events with four-top quark final states, it is currently possible to fit the $\sigma(pp \to t\bar{t}t\bar{t})$ excess in ATLAS-CONF-2020-013 without running afoul of ATLAS-CONF-2020-002 in most cases.

\begin{figure}\label{f10}
\centering
\includegraphics[scale=0.78]{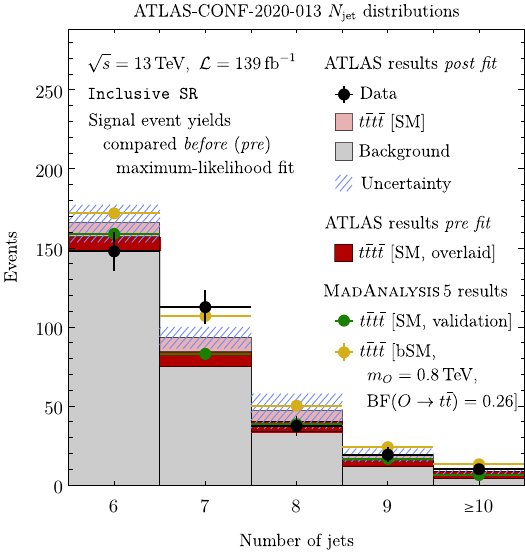}\hspace{1cm}\includegraphics[scale=0.78]{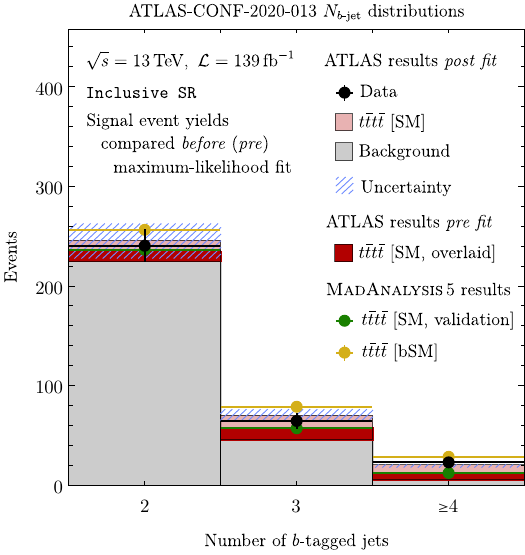}
\caption{Comparison between data and predictions for inclusive signal region events for the distributions of (left) $N_{\text{jet}}$ and (right) $N_{b\text{-jet}}$. Gray background results were obtained (and simplified) from Figure 6 of ATLAS-CONF-2020-013 \cite{ATLAS:20204t} and apply after the maximum-likelihood fit. Light (dark) red results were likewise obtained post (pre) fit from ATLAS and are overlaid, not stacked. Green results are derived from \textsc{MadAnalysis}\,5 for SM $t\bar{t}t\bar{t}$ production. Gold results are derived from \textsc{MadAnalysis}\,5 for $t\bar{t}t\bar{t}$ events mediated by a pair of $800\,\text{GeV}$ scalar sgluons with branching fraction appropriate to fit thirty-event excess. Our sgluon yields are combined with ATLAS backgrounds and ATLAS pre-fit $t\bar{t}t\bar{t}$ yields.}
\end{figure}

One point in parameter space ostensibly particularly well suited to fit the excess is at $m_O\ \text{or}\ m_o = 800\,\text{GeV}$ and $\text{BF}(O\,\text{or}\,o \to t\bar{t}) \approx 0.26$. This point is quite close to both the central value of the excess, providing about thirty extra events, and to the curve corresponding to the $t\bar{t}$ branching fraction predicted for a CP-even sgluon in the minimal $R$-symmetric model depicted in \hyperref[f3]{Figure 3}. This benchmark point offers us an opportunity to take a closer look at the purported excess and probe sgluon-mediated contributions to the four-top quark cross section at a level deeper than the total yields. This is possible on a technical level because (a) the \textsc{MadAnalysis}\,5 framework supports declaration of histograms for any observable on which a selection cut can be imposed, and (b) because ATLAS-CONF-2020-013 reported the predicted and observed distributions of several relevant observables for events in the inclusive signal region after the maximum-likelihood fit. We reproduce the histograms for the numbers of jets, $N_{\text{jet}}$, and $b$-tagged jets, $N_{b\text{-jet}}$, in the left and right panels of \hyperref[f10]{Figure 10}. An excess is particularly visible in the $N_{\text{jet}} = 7$ bin of the left panel, but also exists in the $N_{b\text{-jet}} \geq 4$ (overflow) bin of the right panel. For visual simplicity, we reproduce only the total backgrounds; the individual backgrounds (large enough to be seen at this scale) are provided in Figure 6 of ATLAS-CONF-2020-013. Another difference between our figures and those of ATLAS is our handling of the signal events: we reproduce ATLAS' post-fit $t\bar{t}t\bar{t}$ yields in light red and \emph{superimpose} upon those, in dark red, the pre-fit yields, which are derived by rescaling the post-fit yields according to the signal strength $\mu \approx 2.0$. We perform this rescaling for two reasons. First, we see an opportunity to further validate our \textsc{MadAnalysis}\,5 reimplementation of ATLAS-CONF-2020-013 by comparing the jet distributions of our SM $t\bar{t}t\bar{t}$ sample and the ATLAS pre-fit signal prediction. We indeed find good agreement after normalizing our \textsc{MadAnalysis}\,5 distributions to the total yield of 28.4 events reported in \hyperref[t4]{Table 4}. Second, we are specifically exploring whether a total $t\bar{t}t\bar{t}$ yield consistent with the sum of the SM prediction and a contribution mediated by color-octet scalars can fit the shape of the excess. We therefore combine the yield in each bin for a sample of pair-produced $m_O = 800\,\text{GeV}$ scalar sgluons, with a total yield of 29.6 events for $\text{BF}(O \to t\bar{t}) = 0.26$, with the pre-fit SM $t\bar{t}t\bar{t}$ yields, and compare these to the data, assuming the same backgrounds for simplicity. We notably find that the combined signal yield is consistent, within uncertainty, with the significant excess in the $N_{\text{jet}} = 7$ bin. It should be noted that the background yields are still post fit, but only the $t\bar{t}W + \text{jets}$ background is given a normalization factor greater than unity as part of the maximum-likelihood fit \cite{ATLAS:20204t}. A more careful analysis of $t\bar{t}W + \text{jets}$ events in models with Dirac gauginos may be warranted, since such signals can be generated by a pair-produced electrically neutral $\mathrm{SU}(2)_{\text{L}}$ adjoint (isospin-triplet) scalar \cite{Carpenter:2021EW}. On balance, however, this rough analysis further suggests that a sgluon pair decaying to $t\bar{t}t\bar{t}$ may produce a signal compatible with the mild excess reported by ATLAS. 

\subsection{Looking ahead to the high-luminosity LHC}
\label{s4.2}

\begin{figure}\label{f11}
\centering
\includegraphics[scale=0.7]{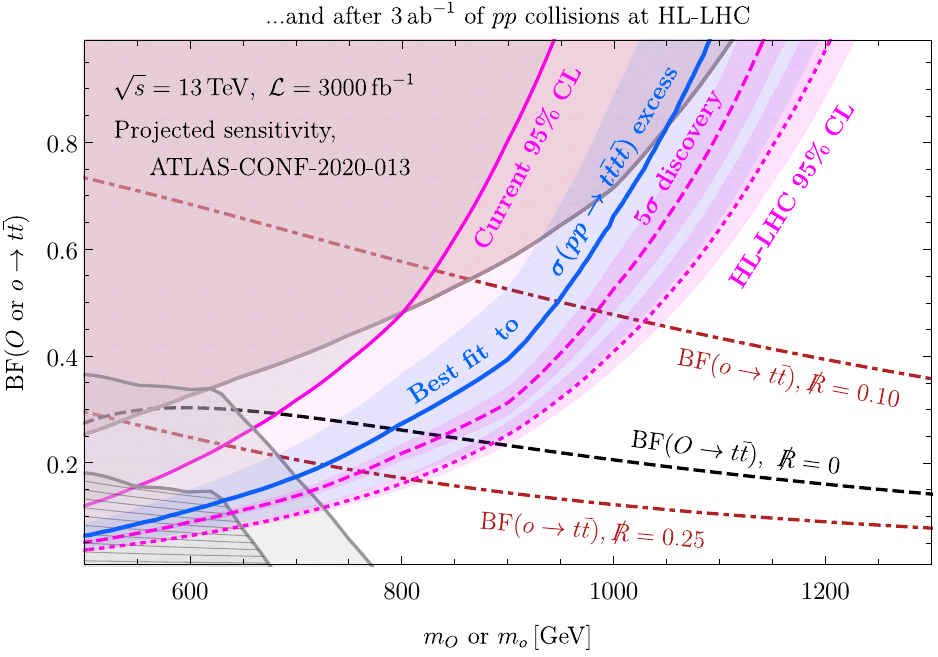}\\[3ex]
\includegraphics[scale=0.7]{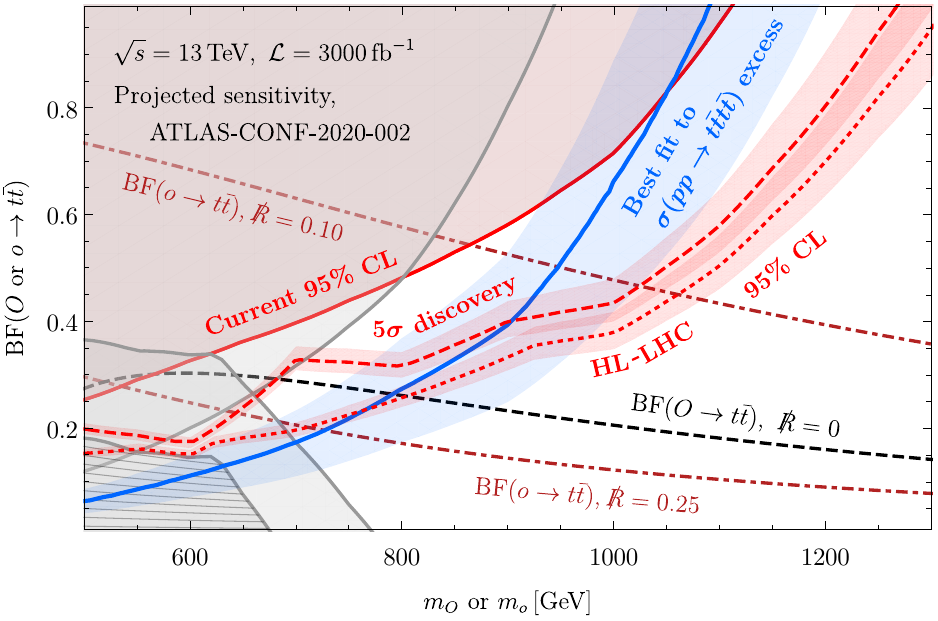}
\caption{Maps of color-octet scalar parameter space, based on \hyperref[f9]{Figure 9}, showing projected sensitivity of upgraded high-luminosity Large Hadron Collider (HL-LHC) given planned integrated luminosity $\mathcal{L} = 3\, \text{ab}^{-1}$. Solid contours are carried over from \hyperref[f9]{Figure 9} and show current limits at 95\% CL for comparison. Dashed contours show projected $5\sigma$ (discovery) signal significance for each analysis. Dotted contours show projected exclusion limits at 95\% CL. Shaded bands around projection contours indicate uncertainties due to PDF and scale variations.}
\end{figure}

Before we conclude, it is worthwhile to consider how durable these results will be against the high-luminosity upgraded LHC (HL-LHC) scheduled to begin collecting data by the end of this decade. To this end, we provide in Figure \hyperref[f11]{11} some views of the same color-octet scalar parameter space mapped in \hyperref[f9]{Figure 9} fast-forwarded to reflect the full planned integrated luminosity of the LHC, $\mathcal{L} = 3\,\text{ab}^{-1}$.
This figure shows projected exclusion limits at 95\% CL for ATLAS-CONF-2020-013 (in the upper panel) and ATLAS-CONF-2020-002 (in the lower) after $3\,\text{ab}^{-1}$ of $pp$ collisions, assuming (for ATLAS-CONF-2020-013) the excess over Standard Model expectation disappears or (for ATLAS-CONF-2020-002) no excess appears. The figure also shows for each analysis a projected contour of signal significance corresponding to the threshold for discovery at five standard deviations relative to background. We estimate the high-luminosity signal significance for each analysis according to
\begin{align}\label{sig}
\mathcal{S}(s_{\text{HL-LHC}}) = \frac{s_{\text{HL-LHC}}}{\sqrt{b_{\text{HL-LHC}}}},
\end{align}
where $s_{\text{HL-LHC}}$ and $b_{\text{HL-LHC}}$ are the numbers of signal and background events expected to survive the selection cuts after the HL-LHC achieves its planned total integrated luminosity. The expression \eqref{sig} yields a pure number, which we interpret as the number of standard deviations at which the signal $s_{\text{HL-LHC}}$ can be discovered. In the interest of simplicity, we neglect the planned increase to a center-of-mass energy of $\sqrt{s} = 14\, \text{TeV}$ for these projections. We also ignore planned upgrades to the detector (which generically ought to improve signal sensitivity) and e.g. pile-up effects due to increased luminosity (which could degrade it). We therefore simply scale up the Run 2 signal and background yields by a factor of the increased luminosity. We compute the signal significance for each analysis using the central values of the sgluon pair-production cross sections displayed in \hyperref[f6]{Figure 6}, but for these we also display the uncertainty bands due to parton distribution function and scale variations. It is clear from \eqref{sig} that we compute the most optimistic possible signal significances by ignoring uncertainties in the HL-LHC background yields. On the other hand, our projected exclusion limits at 95\% CL do include estimates of the future background uncertainties: these are estimated by rescaling the ATLAS Run 2 uncertainties by a factor of the square root of the luminosity increase \cite{Araz_2020}. The resulting limits are fairly aggressive but still somewhat realistic.

There are several features worth highlighting of the two plots in \hyperref[f11]{Figure 11}.  We see that if an excess in four-top events persists, the entire line fitting the central value of the present-day excess will be discoverable at 5$\sigma$ by the current four-top search. This search is uniquely suited to find low-mass sgluons: for example, an $R$-symmetric scalar sgluon lighter than $m_O \approx 850\,\text{GeV}$ could be discovered only in this channel. On the other hand, sgluons heavier than about $900\,\text{GeV}$ will have complementary discovery signatures able to be probed by the jets + $E_{\text{T}}^{\text{miss}}$ search for new phenomena. In fact, this search has the power to discover heavier sgluons in a sizable region of parameter space beyond the current best-fit region. In particular, a pseudoscalar sgluon in an $R$-symmetric scenario is discoverable up to $m_o \approx 1.2\,\text{TeV}$. If a significant excess in either channel is discovered, the complementarity of these two ATLAS analyses will be key to separating hypotheses for new physics explanations. By comparing the two measurements we could, for example, discriminate a low-mass signal (which would appear only in the four-top channel) from a high-mass signal (which would appear only in the multijet channel) and an intermediate-mass signal (which would appear in both channels). A signal appearing in only one channel would also likely imply something important about the color-octet scalar that produced it. A signal from a heavy sgluon in the multijet channel, for instance, would almost certainly indicate a pseudoscalar sgluon with minimal $R$ symmetry breaking (or a scalar sgluon in an $R$-symmetric scenario with disconcertingly heavy squarks). By contrast, a signal from a light sgluon in the four-top channel would suggest either an $R$-symmetric scalar sgluon accompanied by relatively naturally light squarks and (Dirac) gluino or a pseudoscalar with increasingly significant $R$ symmetry breaking. In this latter scenario, and in cases where a signal of moderate mass is discovered in both channels, it would behoove us to look for corresponding signals in diboson channels ($gg$, along the lines of ATLAS-SUSY-2016-09, and $g\gamma$) to disentangle the surviving hypotheses.

On the other hand, we predict that the increased luminosity of HL-LHC will render the entire best-fit region excludable at 95\% CL if no further excess is observed. ATLAS-CONF-2020-013 could by itself strengthen the current limit on pseudoscalar sgluons in minimal $R$-symmetric models to $m_o \gtrsim 1.2\, \text{TeV}$. Interestingly, we see a similar picture for ATLAS-CONF-2020-002, which --- recall --- already excludes a small part of best-fit parameter space at 95\% CL. In particular, we predict that this analysis will be able to exclude all best-fit parameter space for sgluons decaying to top quarks with branching fraction larger than about 25\%. These plots remind us once more that the search for new phenomena is more sensitive to heavy sgluons than the $\sigma(pp \to t\bar{t}t\bar{t})$ measurement: to wit, ATLAS-CONF-2020-002 can exclude minimal $R$-symmetric pseudoscalar sgluons lighter than $m_o \approx 1.4\, \text{TeV}$, a potential improvement of $200\,\text{GeV}$ upon the other analysis. All told, we find that the entire color-octet scalar parameter space well suited to fit the $\sigma(pp \to t\bar{t}t\bar{t})$ excess can be excluded at 95\% CL by ATLAS-CONF-2020-013, and much of it can be excluded twice over by ATLAS-CONF-2020-002. We further find that sizable regions of parameter space unrelated to the four-top excess can be excluded by each analysis during the run of the HL-LHC. In the absence of a discovery, these two analyses would still leave open space for heavy ($\gtrsim\! 1.1\, \text{TeV}$) sgluons with moderate $t\bar{t}$ branching fractions. A pseudoscalar sgluon consistent with the benchmark characterized by $\slashed{R}=0.25$ considered in \cite{Carpenter2021coloroctet} would appear to remain viable, as would a scalar sgluon in a minimal $R$-symmetric model whose gluino is heavy enough relative to the squarks to suppress $\text{BF}(O \to t\bar{t})$. It should be noted, however, that searches like ATLAS-SUSY-2016-09 that constrain color-octet scalars decaying to gluons will also expand their reach into this region. At any rate, after the full scheduled run of the LHC, analyses like those we have studied here should be able to help us discern with greater certitude whether color-octet scalars in models with Dirac gauginos can explain an excess in events with four top quarks.
\section{Conclusions}
\label{s5}

In this paper, we have explored the ability of the color-octet scalars (sgluons) in models with Dirac gauginos, of the variety studied in \cite{Choi:2009co,Benakli:2013mdg,Benakli:2014cmdg,Chalons:2019md,Carpenter:2020mrsm,Carpenter2021coloroctet}, to enhance the observed cross section of production of four top quarks $\sigma(pp \to t\bar{t}t\bar{t})$ at the LHC. This survey was motivated by measurements of this cross section, announced not long ago by the ATLAS collaboration, that deviate from the state-of-the-art Standard Model prediction by a factor of two, or by almost 2.0 standard deviations. This measurement is interesting in light of similar measurements by both LHC collaborations and other searches for new phenomena that find no excess in events with four top quarks. In order to better understand these new results, we have reimplemented the four-top search in multilepton final states using the \textsc{MadAnalysis\,5} framework, alongside an ATLAS search for new phenomena in final states with large jet multiplicities and missing transverse energy, and we have reinterpreted both analyses in the context of the aforementioned color-octet scalar models.

Our survey has yielded some interesting results. In particular, we have found that the two analyses we have reimplemented exhibit some complementarity: the four-top cross section measurement boasts higher sensitivity to our model signals for lighter sgluons, while the search for new phenomena takes precedence as the sgluons decaying to $t\bar{t}$ approach the TeV scale. Altogether, we have found that either species of sgluon in these models can mediate a contribution to the four-top cross section in scenarios with TeV-scale sgluons, low multi-TeV stop squarks and Dirac or pseudo-Dirac gluinos. Specific scenarios that appear promising include a scalar sgluon with unbroken $R$ symmetry and heavy stop squarks and a pseudoscalar sgluon with mildly broken $R$ symmetry. On the other hand, a pseudoscalar in minimal $R$-symmetric models capable of fitting the central value of the excess is ruled out. Taken together, the ATLAS analyses we have considered disfavor color-octet scalars or pseudoscalars in scenarios with masses between $500\,\text{GeV}$ and $650\,\text{GeV}$ and low $t\bar{t}$ branching fractions and exclude pseudoscalars in $R$-symmetric models with masses between $m_o \approx 350\,\text{GeV}$ and $m_o \approx 1.1\,\text{TeV}$. 

Looking forward, we have found that in the full planned dataset of $3\,\text{ab}^{-1}$ for the HL-LHC, the measurement of $\sigma(pp \to t\bar{t}t\bar{t})$ has a $5\sigma$ discovery potential that covers a large swath of color-octet scalar parameter space, extending to about $1.15\,\text{TeV}$. Meanwhile, the jets + $E_{\text{T}}^{\text{miss}}$ search for new phenomena provides an alternative discovery channel with sensitivity past $1.2\,\text{TeV}$ for color-octet scalars with maximal $t\bar{t}$ branching fractions. The existence of not one but two discovery channels with comparable sensitivity (depending on region of parameter space) is fortuitous. If a new particle is discovered in one of these channels, we expect a complementary signal in the other; the existence of two channels offers a key validation mechanism for discovery. On the other hand, large regions of our parameter space may be ruled out, including the region currently fitting the four-top excess, if no further excess of events are measured in either search.

New multitop analyses will be key to further study of sgluon phenomenology. As we have seen, the future $\sigma(pp \to t\bar{t}t\bar{t})$ measurements and jets + $E_{\text{T}}^{\text{miss}}$ searches will be pivotal to hypothesis discrimination if a significant excess evolves in either or both channels. On the other hand, HL-LHC searches in alternate channels may be needed for further hypothesis differentiation or exclusion of color-octet scalar parameter space. Most important are the diboson searches, especially for resonances decaying to gluon pairs. For example, an excess in the $gg$ channel concurrent with one in the four-top channel would point specifically to a light sgluon --- either an $R$-symmetric scalar or a pseudoscalar with moderate $R$ symmetry breaking --- while the absence of this signal could rule out a light sgluon of either kind. In addition, in this same region, decays of color-octet scalars to $g\gamma$ and $gZ$ are non-negligible and can further discriminate between scalar and pseudoscalar states \cite{Carpenter:2015gua}. The latter channel may be particularly relevant, since pair-produced sgluons each undergoing a different decay could produce $t\bar{t}Z + \text{jet(s)}$, itself a significant background in the four-top quark production analysis considered in this work. Sensitivity studies in these channels for the high-luminosity run of the LHC would therefore be one useful avenue of future study.


\acknowledgments
This research was supported in part by the United States Department of Energy under grant DE-SC0011726. We are grateful to C\'{e}line Degrande for technical assistance with \textsc{FeynRules} and \textsc{NloCT}. We are in debt to several of the many excellent analyses \cite{Darme:2018rec,Ambrogi_rec,Araz_2021} on the \textsc{MadAnalysis\,5} Public Analysis Database (PAD) \cite{Dumont_2015}, which we used as guides while developing our own reimplementations.

\bibliographystyle{Packages/JHEP}
\bibliography{Bibliography/bibliography.bib}

\end{document}